\documentclass[aps,prl,superscriptaddress,twocolumn,amsmath,longbibliography]{revtex4-2}

\usepackage{graphicx, color}
\usepackage{xcolor}
\usepackage{amsfonts,amssymb,amsmath}
\usepackage{siunitx}
\usepackage{bm}
\usepackage{bbold}
\usepackage{braket}
\usepackage{mathtools}
\usepackage{dsfont}
\usepackage{lineno}
\usepackage{bigints}
\usepackage[normalem]{ulem}

\newcommand{\NEW}[1]{{}{\textcolor{black}{ #1}}}

\begin{document}

\title{Nonlinear transport due to magnetic-field-induced flat bands in the nodal-line semimetal ZrTe$_5$}

\author{Yongjian Wang}
\affiliation{Physics Institute II, University of Cologne, Z\"ulpicher Stra{\ss}e. 77, 50937 K\"oln, Germany}

\author{Thomas B\"omerich}
\affiliation{Institute for Theoretical Physics, University of Cologne, Z\"ulpicher Stra{\ss}e 77, 50937 K\"oln, Germany}

\author{Jinhong Park}
\affiliation{Institute for Theoretical Physics, University of Cologne, Z\"ulpicher Str. 77, 50937 K\"oln, Germany}

\author{Henry~F.~Legg}
\affiliation{Institute for Theoretical Physics, University of Cologne, Z\"ulpicher Str. 77, 50937 K\"oln, Germany}
\affiliation{Department of Physics, University of Basel, Klingelbergstrasse 82, CH-4056 Basel, Switzerland}

\author{A. A. Taskin}
\affiliation{Physics Institute II, University of Cologne, Z\"ulpicher Str. 77, 50937 K\"oln, Germany}

\author{Achim Rosch}
\affiliation{Institute for Theoretical Physics, University of Cologne, Z\"ulpicher Str. 77, 50937 K\"oln, Germany}

\author{Yoichi Ando}
\email[]{ando@ph2.uni-koeln.de}
\affiliation{Physics Institute II, University of Cologne, Z\"ulpicher Str. 77, 50937 K\"oln, Germany}

\begin{abstract}
The Dirac material ZrTe$_5$ at very low carrier density was recently found to be a nodal-line semimetal, where ultra-flat bands are expected to emerge in magnetic fields parallel to the nodal-line plane. Here we report that in very low carrier-density samples of ZrTe$_5$, when the current and the magnetic field are both along the crystallographic $a$ axis, the current-voltage characteristics presents a pronounced nonlinearity which tends to saturate in the ultra quantum limit. The magnetic-field dependence of the nonlinear coefficient is well explained by the Boltzmann theory for flat-band  transport, and we argue that this nonlinear transport is likely due to the combined effect of flat bands and charge puddles, the latter appear due to very low carrier densities. 
\end{abstract}

\maketitle
\newpage

%\section*{Introduction}

In a nodal-line semimetal (NLSM) \cite{Fang2016}, the crossing of the conduction and valence bands leaves an extended loop of band-touching points, called a nodal line. Together with three-dimensional (3D) Dirac semimetals and Weyl semimetals, NLSMs belong to the family of topological semimetals \cite{Burkov2016, Lv2021} and their low-energy physics is described by a Dirac equation with additional symmetry-breaking terms. 
Intriguingly, it has been theoretically shown \cite{Rhim2015} that in NLSMs, ultra-flat bands generically emerge as a result of Landau quantization in magnetic fields parallel to the nodal-line plane. 
Already a number of materials, including ZrTe$_5$ \cite{Wang2022}, PtSn$_4$ \cite{Wu2016}, PbTaSe$_2$ \cite{Bian2016}, ZrSiSe\cite{Hu2016}, ZrSiTe \cite{Hu2016}, HfSiS \cite{Delft2018}, and CaAgAs \cite{Kwan2020} are found to be NLSMs, but no signatures of flat bands have so far been inferred in their magnetotranport properties.
In this work, we report that in the ZrTe$_5$ samples where the NLSM is realized \cite{Wang2022}, the current-voltage ($I$-$V$) characteristics present an intrinsic nonlinearity, which is greatly enhanced only when the current and the magnetic field are parallel to each other. We argue that the weak nonlinearity in zero magnetic field is most likely due to charge puddles that appear because of very low carrier densities \cite{Wang2022, Skinner2012}, while the pronounced enhancement in the longitudinal magnetic field of a few T is due to flat bands. This scenario is backed up by theoretical calculations based on higher-order Boltzmann theory.

The van-der-Waals material ZrTe$_5$ (see Fig. 1a inset for the crystal structure) is a topological semimetal and has been known for its peculiar temperature-dependent shift of the chemical potential $\mu$ \cite{Weng2014, Xu2018, RYChen2015, Li2016, YZhang2017, Liang2018, HWang2018, Shahi2018, Tang2019, Sun2020, Fu2020, Wang2021}. The temperature dependence of the resistivity in ZrTe$_5$ presents a pronounced peak at a temperature called $T_p$, which, depending on the sample, lies between 0 to 150 K. At $T_p$, the carrier density $n$ shows a minimum and the sign of the charge carriers switches \cite{Xu2018, Tang2019}, because $\mu$ crosses the Dirac point. In samples with $T_p \simeq$ 0 K \cite{Wang2022, Liang2018, Shahi2018} as those used in this work (Fig. 1a), the extremely low $n$ of $\sim$10$^{16}$ cm$^{-3}$ is realized \cite{Wang2022}. 
Such samples were recently found to present a large magnetochiral anisotropy (MCA), which requires broken inversion symmetry \cite{Tokura2018} and points to the change from a 3D Dirac semimetal to a NLSM \cite{Wang2022, Wang2022a}. Indeed, the existence of a torus-shaped Fermi surface, which appears when a NLSM is slightly doped, was confirmed by Shubnikov-de Haas oscillations \cite{Wang2022}. \NEW{Further experimental observations, such as angle-dependent magnetoresistance \cite{Wang2022}, nonlinear Hall effect \cite{Wang2022a}, or anomalous Hall effect \cite{Liang2018}, also support inversion symmetry breaking in low-carrier-density ZrTe$_5$.}

%\section*{Results}

%\subsection{Nonlinear transport}

\begin{figure*}[tb]
	\centering
	\includegraphics[width=13cm]{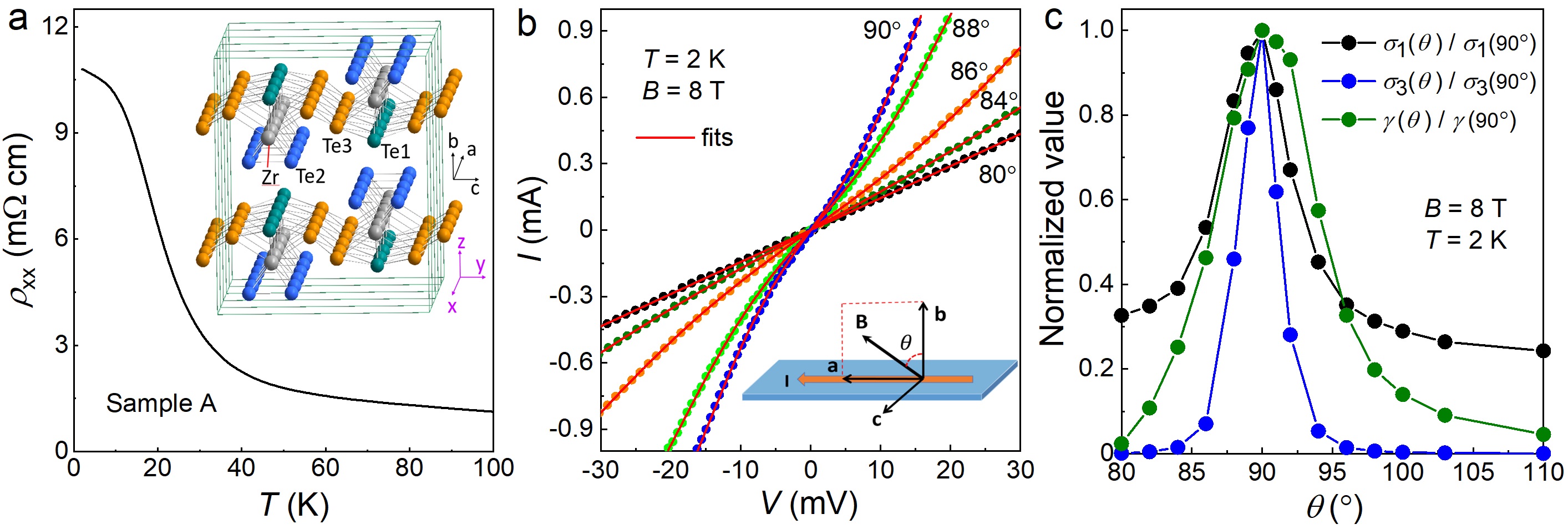}
	\caption{%{\bf Nonlinear longitudinal transport in ZrTe$_5$.} 
{\bf a}, Temperature dependence of $\rho_{xx}$ of sample A with the current $I$ along the $a$ axis. Inset shows the schematic crystal structure of ZrTe$_5$ with ZrTe$_3$ chains along the $a$ axis. 
%Inset shows the schematic crystal structure of ZrTe$_5$ consisting of chains of ZrTe$_3$ running along the $a$ axis and connected through additional Te atoms along the $c$ direction to form a charge-neutral plane, which in turn is stacked along the $b$ axis with van-der-Waals force. 
{\bf b}, $I$-$V$ curves at 2 K and in 8 T for various magnetic-field orientations in the $ab$ plane, measured with DC currents in the $a$-axis. Symbols are the data and solid lines are the fits with $j = \sigma_1 E+\sigma_3 E^3+\sigma_5 E^5$. Inset shows the definition of the magnetic-field orientation $\theta$.
{\bf c}, Dependence of  $\sigma_1$, $\sigma_3$, and $\gamma$ ($\equiv \sigma_3 / \sigma_1^3$) on the magnetic-field angle $\theta$ in the $ab$ plane. The vertical axis is normalized with the maximum values, which are: $\sigma_1^{\rm max}$ = 40.5 $\Omega^{-1} \mathrm{cm^{-1}}$, $\sigma_3^{\rm max}$ = 239.1 $\Omega^{-3} \mathrm{A^{-2} cm}$, and $\gamma^{\rm max}$ = 35.9$\times 10^{-12}$ $\mathrm{m^{4} A^{-2}}$.
}
\label{fig:1}
\end{figure*}

Figure 1b shows the current vs voltage ($I$-$V$) characteristics measured at 2 K in different orientations of the 8-T magnetic field rotated in the $ab$ plane, where the DC current was applied along the $a$-axis; one can see that the $I$-$V$ characteristics become clearly nonlinear when the magnetic field direction is close to the $a$-axis. The existence of MCA (see Supplemental Material \cite{SM}) further evinces the NLSM nature of this sample. Although the data shown in Fig. 1b are measured with a DC current, we have also performed pulsed-current measurements (in which the dissipated power was less than 10\% of that of the DC measurements) and confirmed that the Joule heating is not affecting the data below $\sim$2 mA (see Ref. \cite{SM}); hence, we restrict the DC current measurements to stay below 1 mA, where the nonlinear $I$-$V$ is of intrinsic origin. 
%Details of our samples and the measurement techniques are described in the Methods section. 

To quantify the nonlinearity in the $I$-$V$ characteristics which is clearly antisymmetric with respect to voltage, we fit the curves with the formula $j = \sigma_1 E+\sigma_3 E^3+\sigma_5 E^5$ by calculating the current density $j$ and the electric field $E$ using the sample dimensions; note that $\sigma_1 \propto \tau$ is the linear conductivity with $\tau$ the scattering time, while $\sigma_3 \propto \tau^3$ and $\sigma_5 \propto \tau^5$ characterize the nonlinear components. The solid lines in Fig. 1b are the fits to the data, from which we extracted $\sigma_1$, $\sigma_3$, and $\sigma_5$ (inclusion of higher-order terms only slightly improves the fitting). Since we found that the behavior of $\sigma_5$ is essentially the same as that of $\sigma_3$ (see Ref. \cite{SM}), we focus on $\sigma_3$ for our presentation of the nonlinearity. We note that $\tau$ changes strongly with the magnetic field and its orientation; therefore, to characterize the nonlinearity, it is useful to define the coefficient $\gamma = \sigma_3 / \sigma_1^3$ which is independent of $\tau$. As shown in Fig. 1c, the angular dependence of $\gamma$ shows a sharp maximum at $\theta$ = 90$^{\circ}$, which is the longitudinal transport configuration.

\begin{figure*}[bt]
	\centering
	\includegraphics[width=15.7cm]{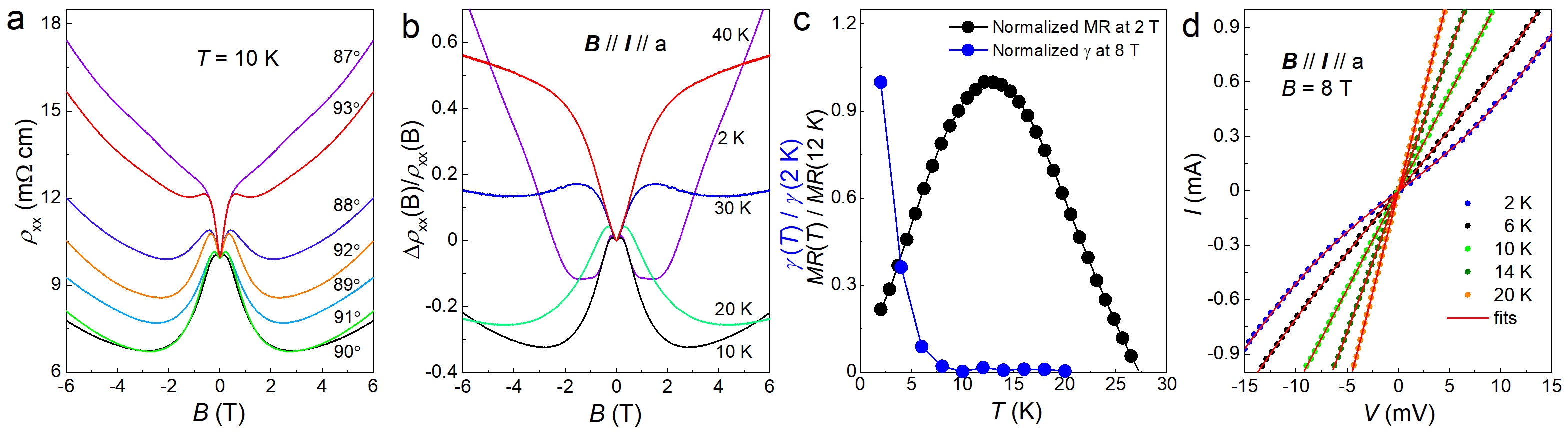}
	\caption{%{\bf Negative longitudinal magnetoresistance (MR) and the nonlinear transport.} 
{\bf a}, MR measured at 10 K with varying magnetic-field orientation $\theta$ near the $a$ axis.
{\bf b}, Longitudinal MR for $B \parallel a$ ($\theta$ = 90$^{\circ}$) measured at various temperatures.
{\bf c}, Comparison of the temperature-dependencies of $\gamma$ (normalized by the value at 2 K) and the negative longitudinal MR at 2 T normalized by the value at 12 K, where it is the largest.
{\bf d}, $I$-$V$ curves for $B \parallel I \parallel a$ measured at various temperatures.
}
\label{fig:2}
\end{figure*}

%\subsection{Role of longitudinal magnetic fields}

In the past, chiral magnetic effect was claimed \cite{Li2016} for ZrTe$_5$ in a sample with $T_p \simeq$ 60 K based on the observation of negative magnetoresistance (MR) which appears only in the longitudinal configuration. The chiral magnetic effect, which is a manifestation of the chiral anomaly, has been studied in Weyl and Dirac semimetals \cite{Ong2021}, but it is not a priori expected in a NLSM that does not have any Weyl nodes. Nevertheless, the negative longitudinal MR has also been observed in $T_p \simeq$ 0 K samples of ZrTe$_5$ \cite{Wang2022}, as is also the case here [Figs. 2a and 2b].
\NEW{Given that there are no Weyl nodes in our samples, this is another example to show that the chiral-anomaly-like feature may be a generic property of a metal near the quantum limit} \cite{Kikugawa2016}. It is useful to mention that the origin of the negative MR seems to have little to do with that of the nonlinearity, because the temperature dependencies of these two effects are very different, even though the configuration is the same: As shown in Fig. 2c, the nonlinear coefficient $\gamma$ diminishes quickly with temperature and vanishes above $\sim$8 K (the $I$-$V$ curves are shown in Fig. 2d), while the negative MR persists to much higher temperature. 
 
In ZrTe$_5$, when the Fermi energy $E_F$ exceeds 20 meV, the Fermi surface takes the usual ellipsoidal shape and the NLSM nature would be lost from the transport properties \cite{Wang2022}. For example, a $T_p$ = 95 K sample with $E_F$ = 25 meV and $n$ = $1.5 \times 10^{17}$ cm$^{-3}$ at low $T$ was reported to have an ellipsoidal Fermi surface \cite{Tang2019}. Indeed, our measurement of a $T_p$ = 133 K sample found no nonlinearity at low $T$ (see Ref. \cite{SM}), suggesting that the very low carrier density and the NLSM nature are crucial for the nonlinear transport in question. %In this regard, the gigantic MCA which signals broken inversion symmetry in ZrTe$_5$ is also a nonlinear transport effect and is limited to the NLSM phase \cite{Wang2022}. Nevertheless, there is a fundamental difference: the MCA is a nonreciprocal property which manifests itself in the second-order response in the transverse configuration, whereas the nonlinear effect reported here is reciprocal and manifests itself in the odd-order (3rd, 5th, ...) response in the longitudinal configuration. 

 \begin{figure}[b]
	\centering
	\includegraphics[width=8.5cm]{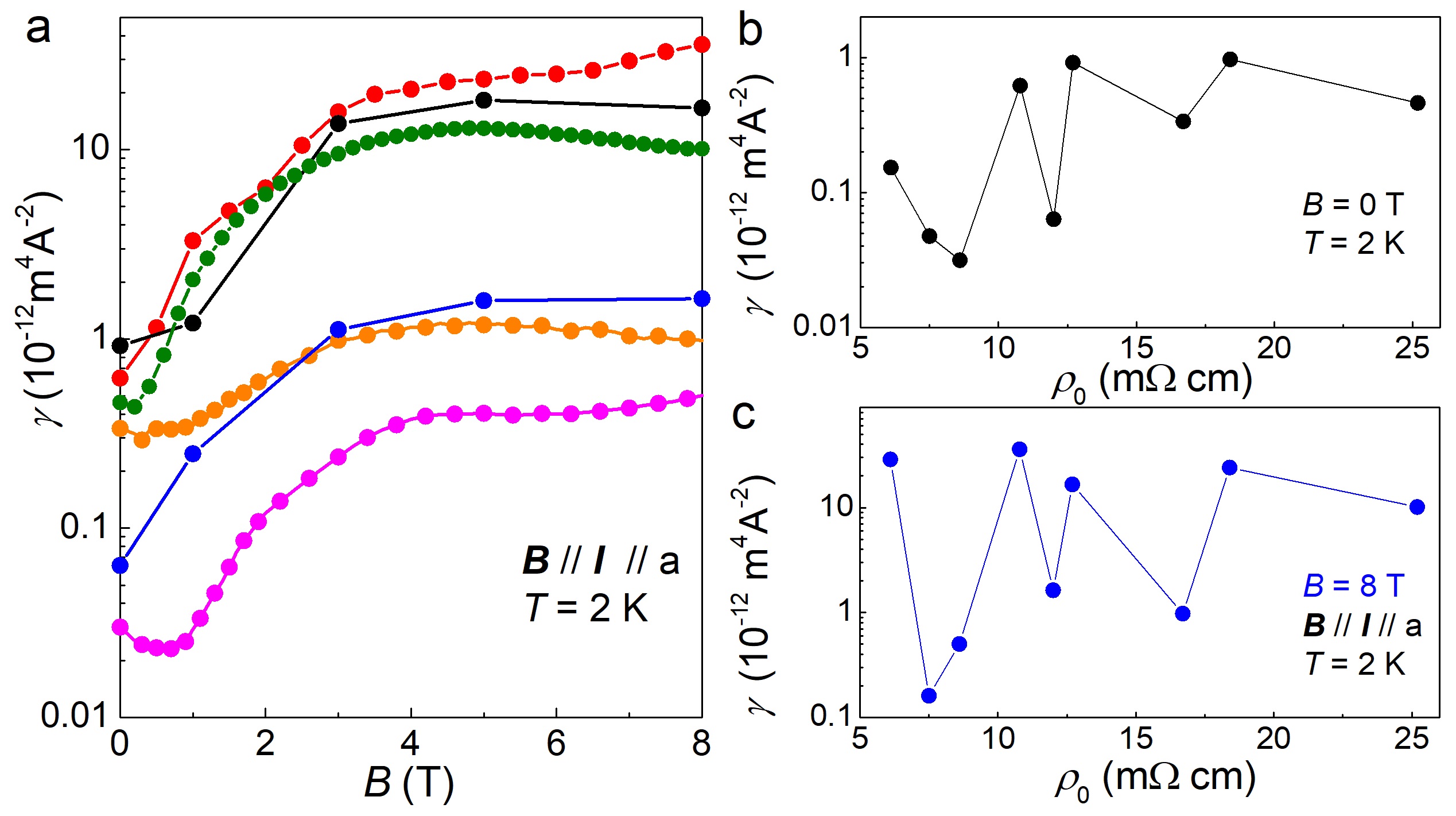}
	\caption{%{\bf Magnetic-field dependence of the nonlinear longitudinal transport.} 
{\bf a}, Magnetic-field dependencies of $\gamma$ measured in various samples (see Ref. \cite{SM} for a summary of samples).
{\bf b-c}, Plots of $\gamma$ vs $\rho_0$ in 0 T (b) and in 8 T (c).
}
\label{fig:3}
\end{figure}

A useful insight into the origin of the nonlinearity is gained by its magnetic-field dependence, which is shown in Fig. 3a for various samples studied. The $\gamma$ value shows a rapid increase with $B$ at a few T, where the system enters the quantum limit \cite{Wang2022}. This behaviour suggests that the nonlinearity develops as the carriers progressively condense into the lowest Landau level upon approaching the quantum limit. 
We note that a nonlinear longitudinal transport associated with the chiral anomaly was previously reported for Bi$_{0.96}$Sb$_{0.04}$ \cite{Shin2017}, in which a gapless 3D Dirac cone is realized. In that case, the nonlinear $\sigma_3$ component was found to show a quadratic magnetic-field dependence (i.e. $\sim B^2$) and persists to temperatures above 20 K, and these features were explained by the Boltzmann transport theory for the chiral anomaly \cite{Shin2017}. Obviously, the nonlinear transport in ZrTe$_5$ does not fall into this category.

%\subsection{Nonlinearity in zero magnetic field and charge puddles}

We note that the nonlinear coefficient $\gamma$ is nonzero even at $B$ = 0 T in Fig. 3a (see Ref. \cite{SM} for $I$-$V$ curves). This means that a small level of nonlinearity already exists in the system without any magnetic field, and this nonlinearity is strongly enhanced with magnetic fields only when the magnetic-field direction is parallel to the current. For the understanding of this zero-field nonlinearity, it is useful to point out that in our experiments, the $\gamma$ value at 0 T is found to be strongly sample-dependent. In Fig. 3b, we plot the zero-field $\gamma$ values measured on 9 samples versus the residual resistivity $\rho_0$. One can see that $\gamma$ varies by more than a factor of 10 and there is no clear correlation between $\gamma$ and $\rho_0$. This strong sample dependence is similar to that found for the magnitude of the MCA, which was argued \cite{Wang2022} to be governed by the details of charge puddles that form inevitably at very low carrier density due to weak screening \cite{Skinner2012}.

Indeed, a recent scanning tunnelling microscope study found that charge puddles appear in ZrTe$_5$ when the carrier density becomes low \cite{Saltzmann2020}. In the presence of puddles, there are various possible mechanisms to cause nonlinear transport in 0 T. For example, when a voltage is applied to the sample, the electric field is screened within metallic puddles and hence is concentrated in the insulating regions between the puddles; this means that the electric field can be locally much larger than the average electric field. In such regions of strong electric fields, the hot-carrier transport, which is the generic mechanism for reciprocal nonlinearity in plain semiconductors \cite{Reggiani1985, Zanette1988}, would take place. Also, the strong local electric field can deform the puddles themselves and eventually lead to percolations, which would also result in nonlinearity \cite{Lippertz2021}. In these puddle-driven scenarios, the strong sample dependence is naturally understood, because how the enhancement of the local electric field occurs will depend on the details of the puddle formation. We leave it to future studies to understand the details of the zero-field nonlinearity, and in the following we focus on the strong enhancement of the nonlinearity in longitudinal magnetic fields (see Fig. 3a). 
As shown in Fig. 3c, the $\gamma$ value in 8 T is also strongly sample dependent and it is larger in samples where the $\gamma$ value is already large in 0 T. \NEW{This suggests that, while the puddle-induced mechanisms cause a large variation in the absolute value of $\gamma$, the magnetic-field-induced nonlinear mechanisms are rather universal.}

%Interestingly, all the experimental observations described above are explained by our theoretical model to consider the formation of a flat band coming from the lowest Landau level in a NLSM and the enhancement of the nonlinearity due to the large-scale inhomogeneity caused by puddles....

%\subsection{Theoretical analysis}

\begin{figure*}[tb]
	\centering
	\includegraphics[width=16.5cm,scale=5.0]{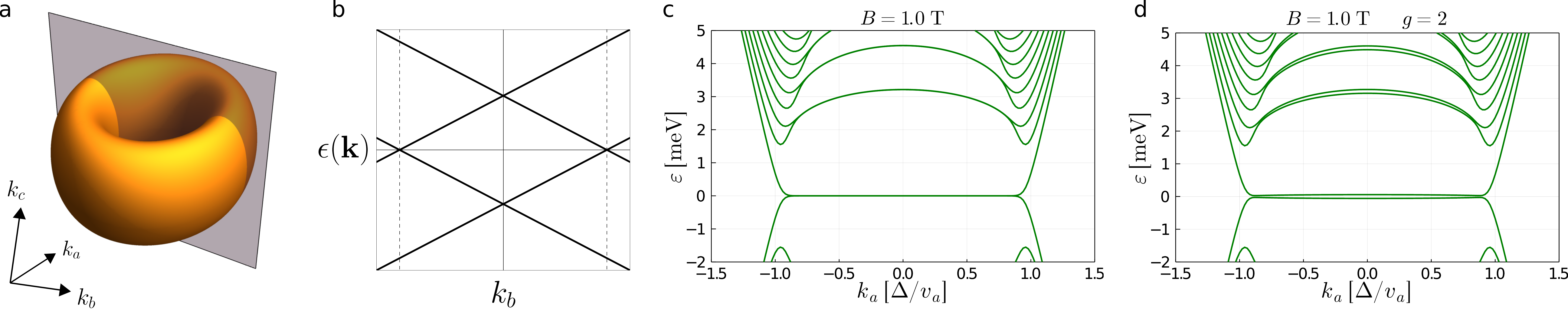}
	\caption{%{\bf Bandstructure and formation of flat band.} 
		{\bf a}, The torus-shaped Fermi surface at small doping, lying in the $ab$ plane. {\bf b}, Band structure for $k_a=k_c=0$. There are two linear crossings which give rise to zero energy states upon Landau quantization. {\bf c}, Landau bands for a magnetic field in the torus plane (see Ref. \cite{SM} for parameters). The lowest Landau band shows an extended region at zero energy. {\bf d}, Including a Zeeman term in the model opens a small gap at zero energy and lifts the degeneracy of higher Landau bands.
	}
	\label{fig:4}
\end{figure*}

Now we show that this magnetic-field-induced enhancement of the nonlinearity is a fingerprint of the formation of the flat bands in NLSMs in the quantum limit \cite{Rhim2015}.
We consider the low-energy Hamiltonian of the NLSM state of ZrTe$_5$ \cite{Wang2022} described in the basis states of $( |\Psi^{\uparrow}_+\rangle, |\Psi^{\uparrow}_-\rangle, |\Psi^{\downarrow}_+\rangle, |\Psi^{\downarrow}_-\rangle)$ with parity $\pm$ and spin $\uparrow/\downarrow$ as follows:
\begin{equation} 
\begin{split} 
H &=m \mathbb{1} \otimes \tau_z+\hbar (v_a k_a \sigma_z \otimes \tau_x +v_b k_b \sigma_x \otimes \tau_x \\ &
+ v_c k_c \mathbb{1} \otimes \tau_y)+\Delta \mathbb{1} \otimes \tau_x  -\mu \mathbb{1}. \label{lowenergyHamiltonian}
\end{split}
%H =m \mathbb{1} \otimes \tau_z+\hbar (v_a k_a \sigma_z \otimes \tau_x +v_b k_b \sigma_x \otimes \tau_x 
%+ v_c k_c \mathbb{1} \otimes \tau_y)+\Delta \mathbb{1} \otimes \tau_x  -\mu \mathbb{1}.  \label{lowenergyHamiltonian}
\end{equation}
Here the Pauli matrices $\sigma_{\alpha}$ and $\tau_{\beta}$ act on the spin and parity space, respectively, and $\Delta$ is the parameter to describe the $ab$-mirror symmetry breaking.
In the minimal Hamiltonian Eq.~\eqref{lowenergyHamiltonian}, upon slight doping ($m < \mu < \Delta$), one obtains a torus Fermi surface lying in the $ab$ plane (Fig.~4a). Since the mass gap $m$ is small at temperatures close to $T_p \sim 0$ K~\cite{Xu2018}, $m \ll \mu$ is justified and the mass term is thus neglected in the following.

When the magnetic field is applied parallel to the torus plane, extremely flat lowest Landau levels emerge along the momenta parallel to the field direction~\cite{Rhim2015}. For a field in the $a$ direction, we consider a cut through the torus Fermi surface at a fixed momentum $k_a$ satisfying $|k_a|<\Delta /v_a $, see Fig.~\ref{fig:4}a. For this fixed $k_a$, the energy dispersion of the Dirac bands for momentum $k_b$ has two Dirac points at $\pm\Delta/v_b$, as shown in Fig.~\ref{fig:4}b. Generally, fixing $k_a$ within the nodal-ring results in a pair of 2D Dirac points in the energy dispersion on the $k_b$-$k_c$ plane.
When the magnetic field is applied, Landau levels develop. Importantly, the lowest Landau level in a 2D Dirac system is always located at zero energy {\em independent} of the value of $k_a$ as long as the field-induced coupling between the two 2D Dirac points can be neglected. Therefore, one obtains two extremely flat bands in the $k_a$ direction parallel to the field and, as the Landau level is also flat in the directions perpendicular to the field, electron motion is strongly suppressed in all directions.
In the absence of Zeeman coupling the velocity of the band is exponentially small (Fig.~\ref{fig:4}c), while the Zeeman effect leads to a small $k_a$-dependent splitting (see Fig.~\ref{fig:4}d and Ref. \cite{SM}).

To calculate nonlinear transport, we use the Boltzmann equation in relaxation time approximation. 
\NEW{The Boltzmann equation is expected to be valid in the limit when scattering rates are much smaller than the Fermi energy and the relevant band gaps, and for not too strong electric fields, $e E \tau v_F \ll \epsilon_F$. For a theory that studied the effects of electric fields of arbitrary strength, we refer to Ref.~\cite{Wang2020}. %For small $E$ also a continuum theory is 
%valid~\cite{Wang2020}, as only small momentum modes are excited.
}
 For our analysis, we use the fact that the nonlinear coefficient $\gamma$ is {\em independent} of the relaxation time $\tau$, which allows to obtain theoretical predictions even if $\tau$ and its field dependence are unknown. In the following, we denote the theoretically-calculated nonlinear coefficient $\gamma_3$, to discriminate it from experimental $\gamma$.
For zero magnetic field, we obtain (see Ref. \cite{SM}) for a torus-shaped Fermi surface characterized by electron density $n$
 \begin{equation}
 \gamma_3(B=0) = \frac{3}{4 e^2 v_a^2} \left( \dfrac{3}{n^2} +16 \pi \frac{\hbar^3 v_a v_b v_c}{\Delta^3 n}\right) \,.
 \label{Gamma3ZeroField}
 \end{equation}
The first term, $\frac{9}{4 e^2 v_a^2 n^2}$, dominates for $n<n_c$ where $n_c$ is the critical density above which the torus Fermi surface does not exist any more. In a pure Dirac system, one obtains a very similar relation, $\gamma_3(B=0) =\frac{6}{5 e^2 v_a^2 n^2}$.
 
 \begin{figure}[b]
	\centering
	\includegraphics[width=6.7cm]{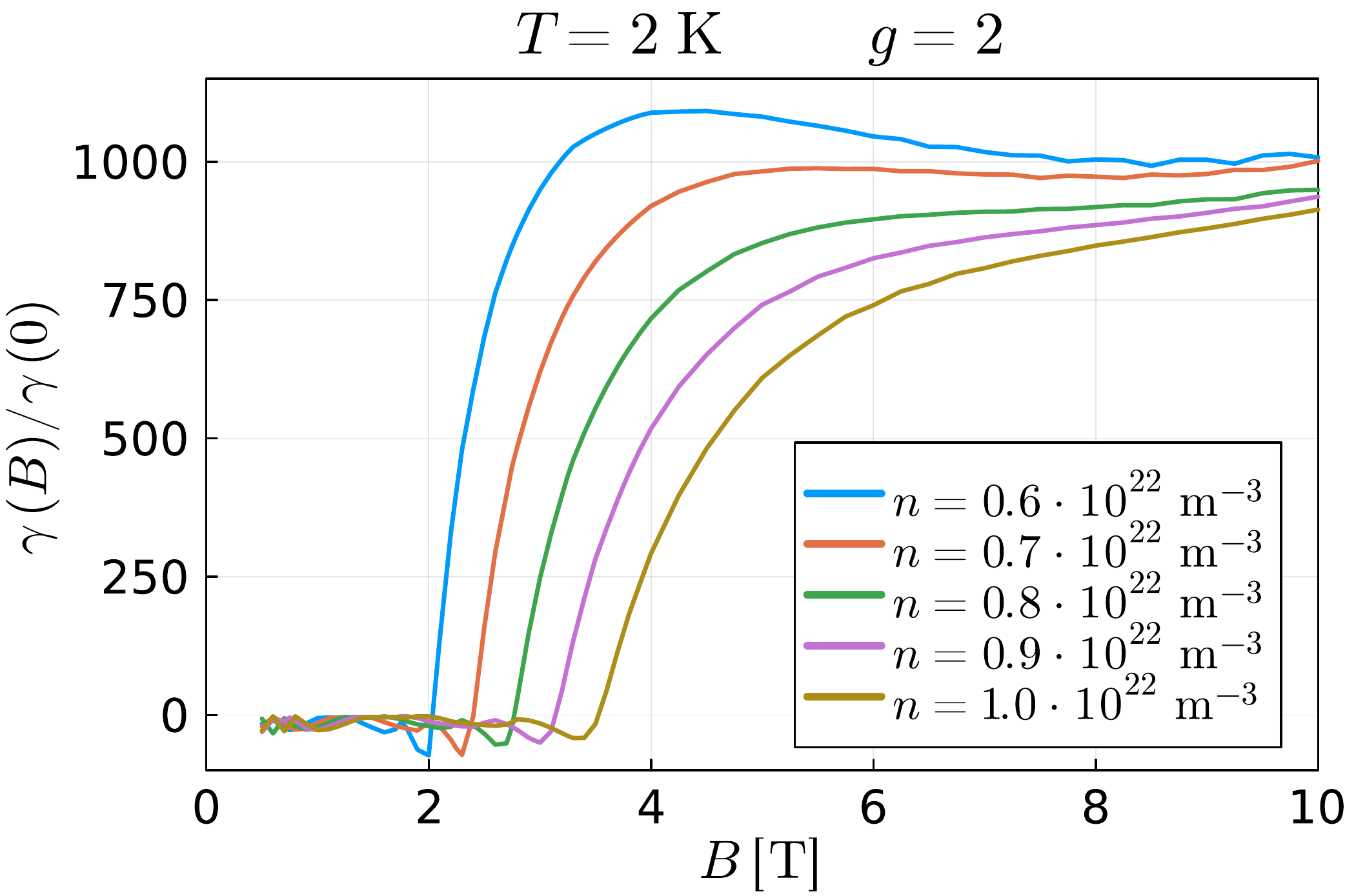}
	\caption{%{\bf Theoretical results for the nonlinearity} 
		{\bf a}, Nonlinear coefficient $\gamma_3$ as a function of magnetic field for different carrier densities. A large enhancement occurs in the flat-band regime. Increasing the field further leads to a saturation of $\gamma_3$. All curves are normalized by the value at 0 T given by Eq. \eqref{Gamma3ZeroField}, \NEW{which are 5.7, 4.2, 3.2, 2.6, and 2.1 ($10^{-18}$ $\mathrm{m^4/A^2}$) for $n$ = 0.6, 0.7, 0.8, 0.9, and 1.0} ($10^{16}$ $\mathrm{cm^{-3}}$), respectively.
	}
	\label{fig:5}
\end{figure}

In the presence of Landau levels only the momentum along the magnetic field is a good quantum number and we obtain multiple 1D Landau bands. We calculate $\gamma_3(B)$ for these bands within the Boltzmann theory (see Ref. \cite{SM} for details) by using the formula
\begin{equation}
\gamma_3(B) = \frac{(2 \pi \hbar)^4}{e^4 B^2} \frac{\sum_n \int \mathrm{d}k_a \frac{\partial \epsilon_{n,k_a}}{\partial k_a} \frac{\partial^3 \epsilon_{n,k_a}}{\partial k_a^3}\frac{\partial f_0}{\partial \epsilon_{n,k_a}} }{\left(\sum_n \int \mathrm{d}k_a \left( \frac{\partial \epsilon_{n,k_a}}{\partial k_a} \right)^2 \frac{\partial f_0}{\partial \epsilon_{n,k_a}} \right)^3} \,.
\label{Gamma3LandauLevels}
\end{equation}
Here, while both numerator and denominator become small in the flat-band limit, the strong dependence on $\left(\partial \epsilon_{n,k_a}/{\partial k_a} \right)^6$ from the denominator dominates. We  assume that the scattering rate $1/\tau$ is small compared to the Landau-level spacing. Figure \ref{fig:5} shows that $\gamma_3(B)$ is strongly enhanced compared to its zero-field value as soon as one reaches the quantum limit where the Fermi energy is located in the flat-band region. As the electron density becomes larger, higher magnetic fields are required to reach the quantum limit. While we obtain $\gamma_3>0$ at $B=0$ and deep in the quantum limit, we also obtain regimes with $\gamma_3<0$ which may, however, be averaged out when a distribution of electron densities is considered.

We note that the magnitude of $\gamma_3(B=0)$ predicted by the Boltzmann theory for a homogeneous system is only $\sim 6\times 10^{-18}\ \mathrm{m^4/A^2}$ for $n = 0.6\times 10^{16}\ \mathrm{cm^{-3}}$, which is several orders of magnitude smaller than what is experimentally observed. As is discussed above, this discrepancy is most likely due to the puddle-induced nonlinear mechanism, the details of which should be clarified in future studies. The magnetic-field induced nonlinear mechanism calculated here provides a pronounced enhancement, $\gamma_3(B)/\gamma_3(0) \sim$ 10$^3$, as shown in Fig.~\ref{fig:5}, and the increase in $\gamma_3(B)$ occurs rapidly at a few T 
\NEW{at the onset of the quantum limit. The magnetic-field dependence qualitatively agrees with the experimental result shown in Fig.~\ref{fig:3}a, but the Boltzmann equation overestimates the amplitude of the enhancement. We attribute this to the fact that the Boltzmann equation is only valid if the scattering rate is smaller than the bandwidth and the relevant band gaps, and thus fails in the flat-band limit, where transport is dominated by impurity-induced inter-band scattering \cite{Abrikosov98}.}
Hence, our theoretical analysis strongly suggests that the pronounced increase of the nonlinear coefficient $\gamma_3$ with $B$ observed experimentally can be explained by the formation of ultra-flat bands in the quantum limit of NLSMs. Further discussions to dismiss trivial effects that might be relevant to nonlinearity, such as current jetting or non-Ohmic contacts, are given in Ref. \cite{SM}.

%Hence, our theoretical analysis strongly suggests that the pronounced increase of the nonlinear coefficient $\gamma_3$ with $B$ observed experimentally can be explained by the formation of ultra-flat bands in the quantum limit of NLSMs. However, the magnitude of $\gamma_3$ at 8 T predicted by the Boltzmann theory for a homogeneous system is $\sim 6\times 10^{-15}\ \mathrm{m^4/A^2}$ for $n\sim 0.6\times 10^{16}\ \mathrm{cm^{-3}}$, which is still orders of magnitude smaller than what is experimentally observed. As is discussed above, this discrepancy is most likely due to the puddle-induced nonlinear mechanism, the details of which should be clarified in future studies. The magnetic-field induced nonlinear mechanism leads to $\gamma_3(B)/\gamma_3(0) \sim$ 10$^3$ as shown in Fig.~\ref{fig:5} and the increase in $\gamma_3(B)$ occurs rapidly at a few T (onset of the quantum limit), which agrees with the experimental result shown in Fig.~\ref{fig:3}a.

%All the $I$-$V$ curves under magnetic field shown in this paper were symmetrized by plus and minus magnetic fields to remove small Hall signal generated by misalignment of the contacts. 

%\section{Discussions}

The ultra-flat bands generically emerge as the lowest Landau level of the Dirac equation to describe a NLSM in magnetic fields applied parallel to the nodal-line plane. Pronounced nonlinear transport in the quantum limit can be understood as a natural consequence of the transport along such a quasi-1D ultra-flat band. The effect of the ultra-flat band can be further enhanced when realised in combination with charge puddles that naturally form in any low carrier density material. The high level of degeneracy in flat bands makes electron correlations to become important, and in fact, the flat band realized in twisted bilayer graphene presents extremely rich physics originating from correlations \cite{Andrei2020}. Hence, it would be interesting to elucidate further consequences of the formation of flat bands in NLSMs, where both topology and correlations may play important roles.
%\cite{Liang2018a, Levin1997, Arnold2016, Zlobin1972, Pinchuk1980, Borgwardt2016, Breunig2017}

\section{Acknowledgements} 
We thank Markus Braden and Jens Brede for useful discussions. This project has received funding from the European Research Council (ERC) under the European Union's Horizon 2020 research and innovation programme (grant agreement No 741121) and was also funded by the Deutsche Forschungsgemeinschaft (DFG, German Research Foundation) under CRC 1238 -- 277146847 (Subprojects A04, B01, and C02) as well as under Germany's Excellence Strategy - Cluster of Excellence Matter and Light for Quantum Computing (ML4Q) EXC 2004/1 -- 390534769. H.F.L. acknowledges funding by the George H. Endress foundation.

\end{document}

{\bf{\large Supplemental Material for ``Nonlinear transport due to magnetic-field-induced flat bands in the nodal-line semimetal ZrTe$_5$''}
}

\vspace{5mm}

{\bf Methods:} 

Single crystals of ZrTe$_5$ with the resistivity-peak temperature $T_{\rm p}$ = 0 K were grown by Te-flux method. The samples with $T_{\rm p}$ = 133 K were grown by a chemical vapor transport method. The details were reported in Ref. \cite{Wang2020}. To make good electrical contacts, the bulk surface was cleaned by Argon plasma to remove the oxidized surface layer, and a thin film of 10 nm Ti / 60 nm Al was deposited on the contact area defined by a mask. All the transport measurements were performed in the four-terminal configuration using a Quantum Design Physical Property Measurement System (PPMS) with a Rotator Probe option. The DC data were measured with Keithley 2450 source meter as DC current source and Keithley 2182 as voltage meter. Pulsed $I$-$V$ curves were measured with Keithley 6221 source meter as a pulsed square-wave generator and a AC lock-in amplifier as a voltage meter; the details of the pulse measurements are described below.

\section{Supplemental Data and Discussions}

\subsection{Summary of the samples measured}

Information on the ZrTe$_5$ single-crystal samples used in this work is summarized in Table I.

\begin{table}[h]
		\caption{Summary of all the samples used for the four-terminal transport measurements. The parameters are: voltage-contact separation $l$, width $w$,  thickness $t$, resistivity-peak temperature $T_{\rm p}$, residual resistivity $\rho_0$ at 2 K, $\gamma$(0 T), $\gamma$(8 T), and the current density $j$(1 mA) at the applied current of 1 mA in the DC $I$-$V$ measurements.}
		\label{table:1}
		{\scriptsize
			\begin{center}
				\begin{tabular}{cccccccccc}
					\hline			
					& Sample \, & $l$ ($\mu$m) \, & $w$ ($\mu$m) \, & $t$ ($\mu$m) \, & $T_{\rm P}$ (K) \, & $\rho_{0}$ (m$\Omega$cm) \, & $\gamma$(0 T) (10$^{-12}$ m$^{4}$A$^{-2}$) \, & $\gamma$(8 T) (10$^{-12}$ m$^{4}$A$^{-2}$) \, & $j$(1 mA) (A/mm$^2$) \, \\
					\hline
					& A & 515 & 115 & 47   & 0   & 10.8 & 0.62  & 35.90  & 185  \\  %0216-S7 (Main text and behavior of \sigma _5)
					& B & 221 & 90  & 14   & 0   & 20.5 & -     & 1.19   & 794  \\  %0917-S1b (Pulsed measurements)
					& C & 218 & 66  & 13.5 & 133 & 0.14 & 0     & 0      & 1122 \\  %S8 (Nonlinear transport in high T_p=133K samples)
					& D & 221 & 90  & 14   & 0  & 16.7  & 0.34  & 0.98   & 794  \\  %0917-S1 (Irrelevance of current jetting effect)
					& E & 314 & 152 & 17   & 0  & 12.7  & 0.92  & 16.6   & 387  \\  %S7 (Reproducibility of the strong angular dependence)
					& F & 280 & 107 & 15.8 & 0  & 12    & 0.06  & 1.63   & 592  \\  %S6
					& G & 220 & 160 & 10   & 0  & 7.5   & 0.05  & 0.16   & 625  \\  %0917-S2
					& H & 220 & 138 & 7    & 0  & 8.6   & 0.03  & 0.5   & 1035 \\  %0918-S3
					& I & 200 & 229 & 20   & 0  & 25.2  & 0.46  & 10.1   & 218  \\  %0810-S2
					& J & 346 & 101.8 & 13 & 0  & 18.4  & 0.97  & 24     & 756  \\  %S1
					& K & 188 & 80 & 42    & 0  & 6.1   & 0.15  & 28.70  & 298  \\  %S2
					\hline
				\end{tabular}
			\end{center}
		}
	\end{table}

\subsection{Pulsed measurements}

\begin{figure}[h]
	\centering
	\includegraphics[width=14 cm]{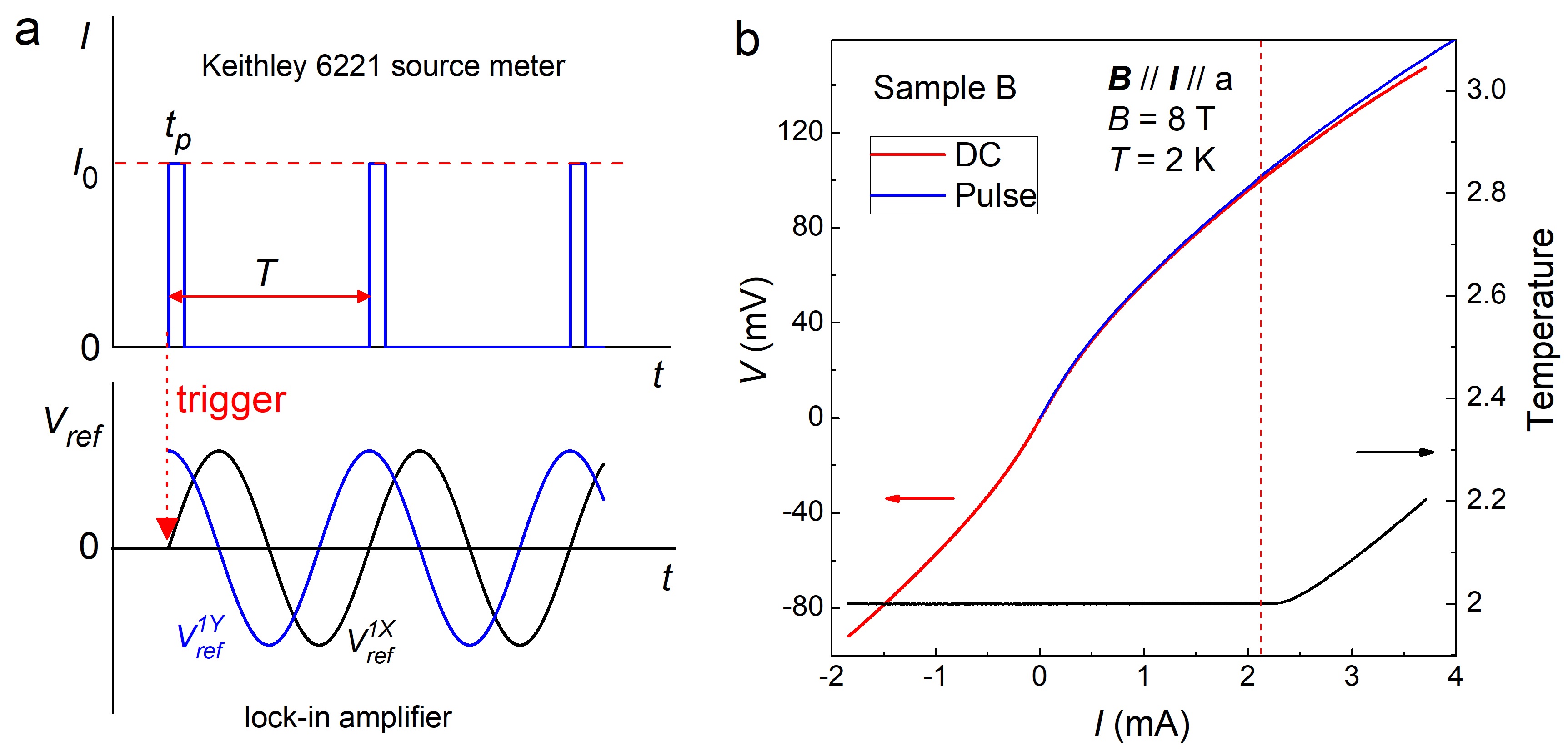}
	\caption{\textbf{Pulsed $I$-$V$ measurement in sample B.} (a) Square wave used for the pulsed measurements (top) and the reference wave form (bottom). (b) Red and blue curves represent the $I$-$V$ curves measured by DC and pulse modes, respectively. Black line shows the current dependence of the temperature measured by a thermometer lcated close to the sample puck inside the PPMS chamber.		
	}
	\label{fig:pulse}
\end{figure}

To elucidate the effect of Joule heating in the $I$-$V$ characteristics, we performed pulsed measurements of $I$-$V$ curves in sample B using a square wave with a frequency of $f$ = 77.7 Hz and duty cycle of 7.77 \%, in which the current switches between 0 and $I_0$ as shown in Fig. \ref{fig:pulse}a. In such a square wave, the width of the current pulse in one period is $t_p$  = 1 ms. The voltage signal was taken by a lock-in amplifier which is triggered by the pulse generator. Defining the temporal resistance at the current $I_0$ to be $R_0$, the out-of-phase component of the AC voltage is given by $V^{1Y}_{\rm out} = \sqrt{2} R_0 I_0 f \int_{0}^{t_p} \cos (2\pi f t) dt $. The parameters of our pulsed measurements give $V_{\rm pulse} \equiv R_0 I_0 = 9.47 V^{1Y}_{\rm out}$. From the reading of the current $I_0$ and voltage $V^{1Y}_{\rm out}$, the pulsed $I$-$V_{\rm pulse}$ curves were obtained, as shown with the blue curve in Fig. \ref{fig:pulse}b. 

If we compare the pulsed and DC measurements at 1 mA in the present case, the Joule heating power is 4.4 $\mu$W for the pulse and 56.8 $\mu$W for DC. Despite this order-of-magnitude difference in the generated heat, there is no difference between the pulsed and DC voltage values for 1 mA, which gives evidence that the Joule heating is not adversely affecting the $I$-$V$ characteristics measured with DC. Actually, for the current values below $\sim$2 mA, there is no visible difference between pulse and DC data, while the Joule heating starts to affect the DC $I$-$V$ curves above $\sim$2 mA. 

During the DC measurements, we also monitored the temperature of the sample holder, which was found to show an increase above $\sim$2.3 mA, as shown with the black line in Fig. \ref{fig:pulse}b. In the pulse mode, there was no temperature change up to 10 mA. Note that the temperature is measured by a thermometer close to the sample puck inside the PPMS chamber, which is kept at a pressure of a few Torr with helium gas to maintain a thermal contact. Since the threshold for the adverse Jouse heating in the DC measurements may slightly vary with samples with different resistivities, we show $I$-$V$ curves for currents below 1 mA for all the samples to guarantee that the observed nonlinearity is intrinsic.

\subsection{Behavior of $\sigma _5$}

\begin{figure}[h]
	\centering
   \includegraphics[width=15 cm]{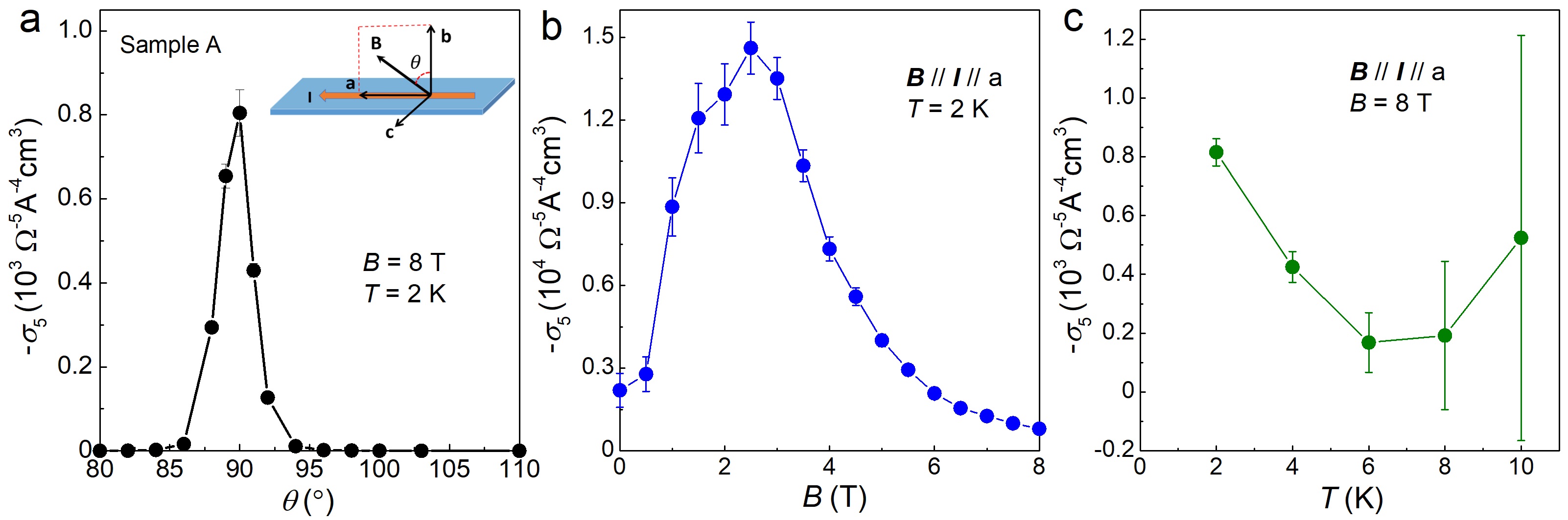}
	\caption{\textbf{Behavior of $\sigma_5$ in sample A.} (a) Angular dependence of $\sigma_5$ obtained from the fits at 8 T and 2 K shown in the main text. (b) Magnetic-field dependence of $\sigma_5$ for $B \parallel a$ at 2 K. (c) Temperature dependence of $\sigma_5$ for $B \parallel a$ at 8 T. Error bars are standard error of the regression.
	}
	\label{fig:sigma5}
\end{figure}

Figure \ref{fig:sigma5} shows the observed behaviour of the coefficient of the 5th-order nonlinearity, $\sigma_5$. It is visible only near $\theta$ = 90$^{\circ}$ (Fig. \ref{fig:sigma5}a), it increases rapidly at 1 -- 2 T as the quantum limit is approached (Fig. \ref{fig:sigma5}b), and it becomes indistinguishable above $\sim$8 K (Fig. \ref{fig:sigma5}c).  This is very similar to the behaviour of $\sigma_3$ presented in the main text.

\vspace{-1cm}

\NEW{
\subsection{Magnetic-field rotation in the $ac$ plane}
\label{sec:ac-rotation}}

\begin{figure*}[h]
	\centering
\NEW{
	\includegraphics[width=10cm]{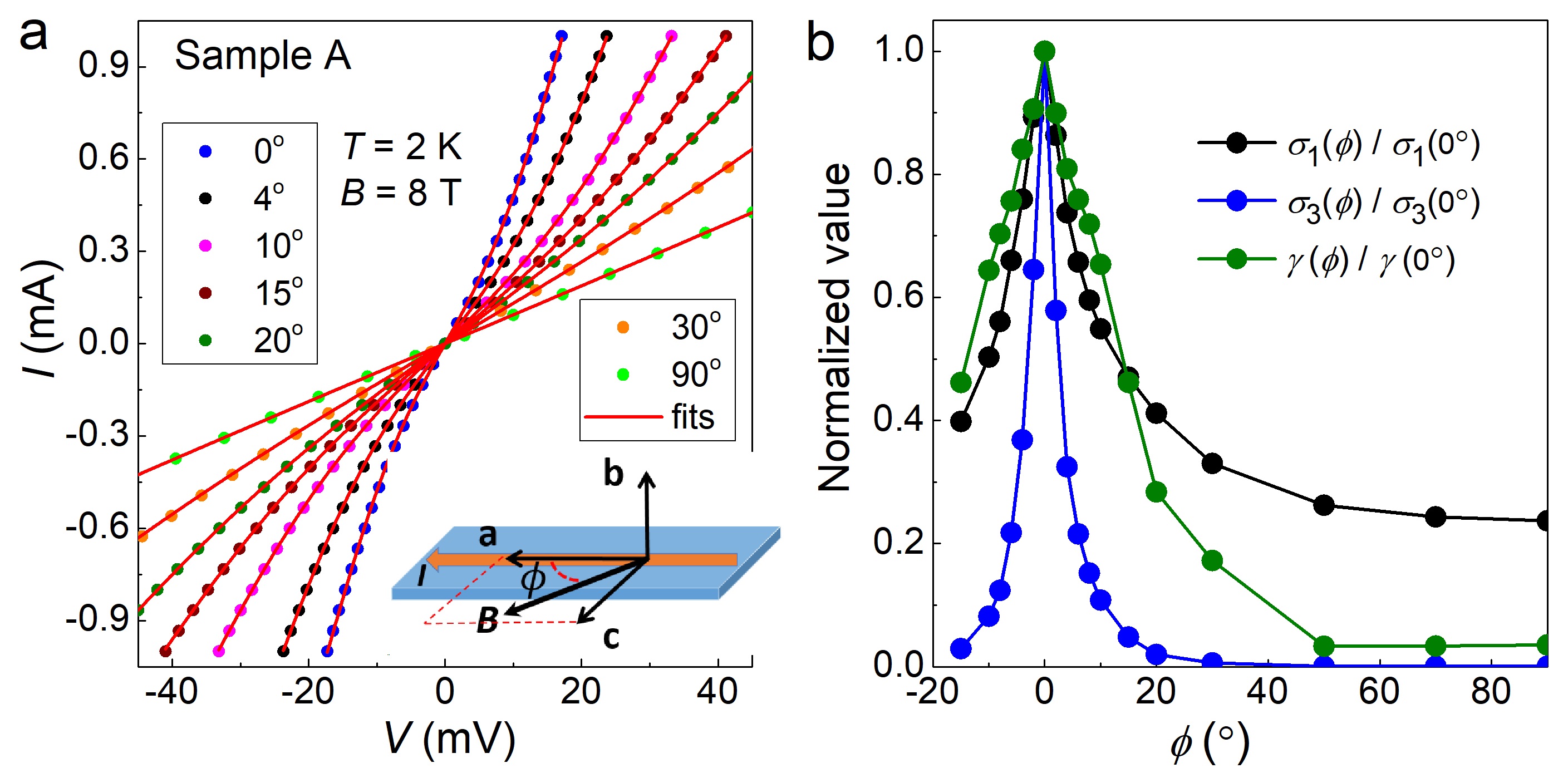}
	\caption{{\bf a}, $I$-$V$ curves at 2 K and in 8 T for various magnetic-field orientations in the $ac$ plane, measured on sample A with DC currents in the $a$-axis. Symbols are the data and solid lines are the fits with $j = \sigma_1 E+\sigma_3 E^3+\sigma_5 E^5$. Inset shows the definition of the magnetic-field angle $\phi$.
{\bf b}, Dependence of  $\sigma_1$, $\sigma_3$, and $\gamma$ ($\equiv \sigma_3 / \sigma_1^3$) on the magnetic-field angle $\phi$ in the $ac$ plane. The vertical axis is normalized with the maximum values at $\phi = 0^{\circ}$.}
  \label{fig:ac-rotation}
  }
\end{figure*}

\NEW{For completeness, we have also measured the magnetic-field-angle dependence of the nonlinear transport in sample A when the magnetic field is rotated in the $ac$ plane. As shown in Fig. \ref{fig:ac-rotation}, the nonlinearity is strongly peaked at the angle $\phi = 0^{\circ}$, which corresponds to the longitudinal configuration, and $\gamma$ decays when $\phi$ moves away from $0^{\circ}$. The decay is slower than the rotation in the $ab$ plane shown in Fig. 1c of the main text, which is presumably due to a difference in the Fermi-velocity anisotropy in the $ac$- and $ab$-planes ($v_a : v_c \simeq 4$, $v_a : v_b \simeq 16$ \cite{Wang2022}), as we explain later in Sec. \ref{sec:field_angle}.
}

\subsection{Magnetochiral anisotropy in sample A}

\begin{figure}[h]
	\centering
   \includegraphics[width=7 cm]{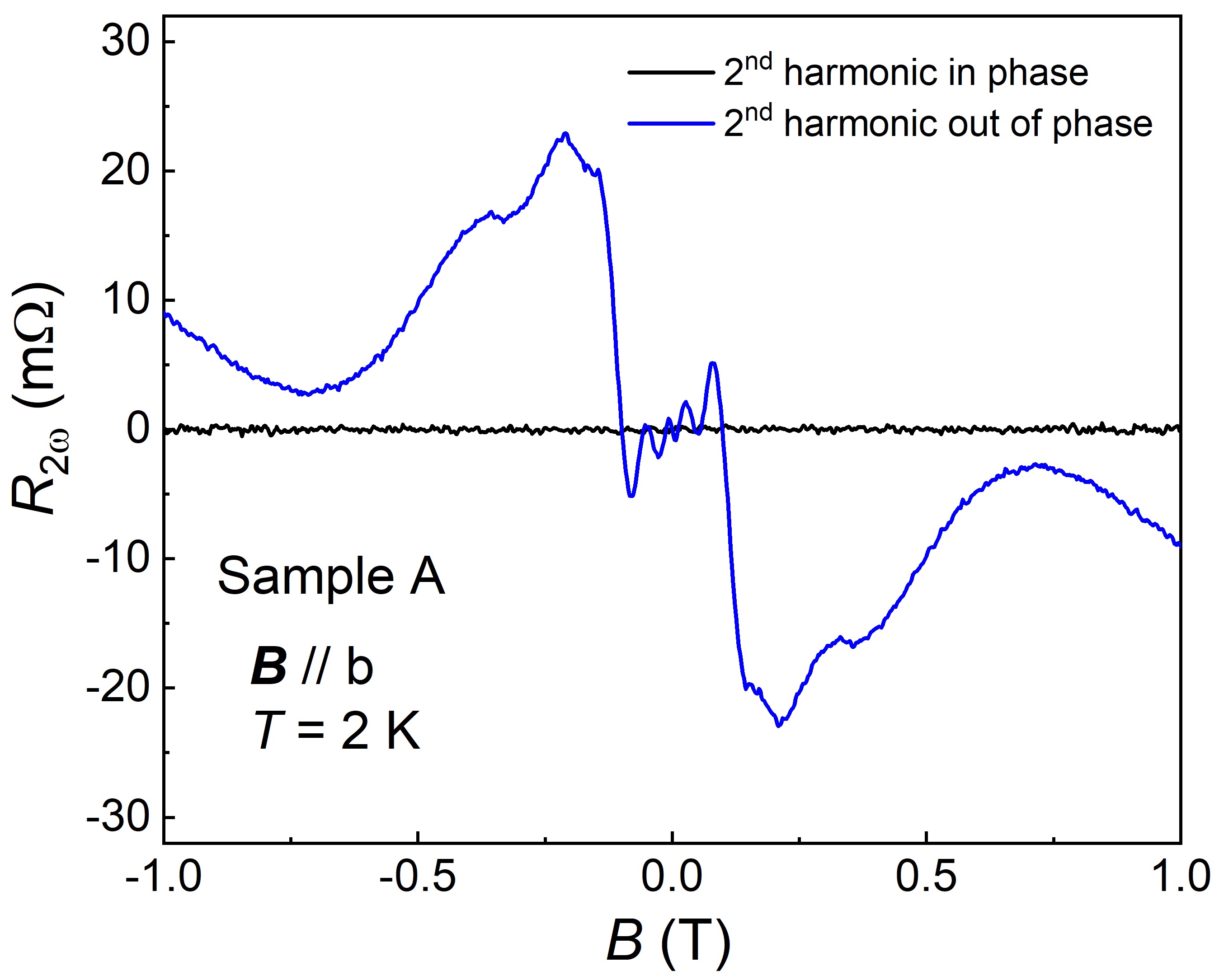}
	\caption{\textbf{Nonreciprocal transport due to torus-shaped Fermi surface.} In-phase and out-of-phase components of the 2nd harmonic resistance $R^{2\omega}$ measured with a low-frequency AC lock-in technique at 2 K as a function of the magnetic field $B$ applied along the $b$-axis. In the presence of the MCA, the out-of-phase signal becomes finite while the in-phase component remain zero.
	}
	\label{fig:MCA-A}
\end{figure}

 In ZrTe$_5$, as mentioned in the main text, the occurrence of a large magnetochiral anisotropy (MCA) is tied to the realization of the NLSM phase and the resulting torus-shaped Fermi surface \cite{Wang2022}. Since the realization of such a Fermi surface in ZrTe$_5$ is dependent on the chemical potential, it is important to confirm the existence of MCA in samples which present the peculiar nonlinear transport described in the main text. Figure \ref{fig:MCA-A} shows the MCA data measured on sample A at 2 K in $B \parallel b$. The 2nd harmonic out-of-phase signal detects the nonreciprocal component due to MCA \cite{Wang2022} and gives confidence in the realization of the NLSM phase.

\subsection{Absence of pronounced nonlinear transport in high $T_p$ samples}

\vspace{-3mm}
\begin{figure}[h]
	\centering
\includegraphics[width=11.2 cm]{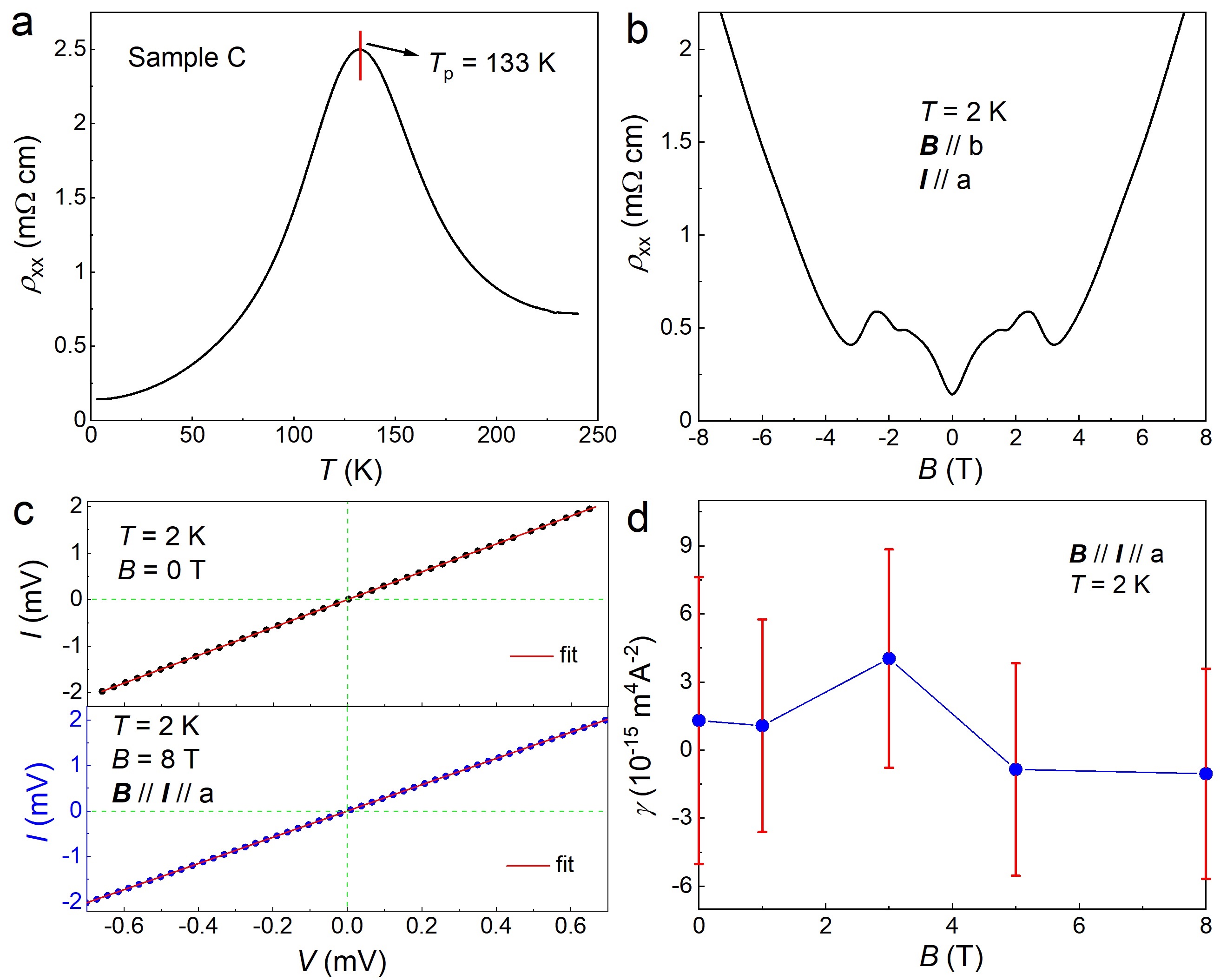}
	\caption{\textbf{Behavior of sample C with $T_p$ = 133 K.} (a, b) Temperature and magnetic field dependencies of the resistivity. (c) $I$-$V$ curves measured at 0 and 8 T with $B \parallel a$ at 2 K. (d) Magnetic-field dependence of $\gamma$ obtained from the fits. Error bars are standard error of the regression.
	}
	\label{fig:HighTp}
\end{figure}

\vspace{-3mm}
\begin{figure}[h]
	\centering
\includegraphics[width=10.2 cm]{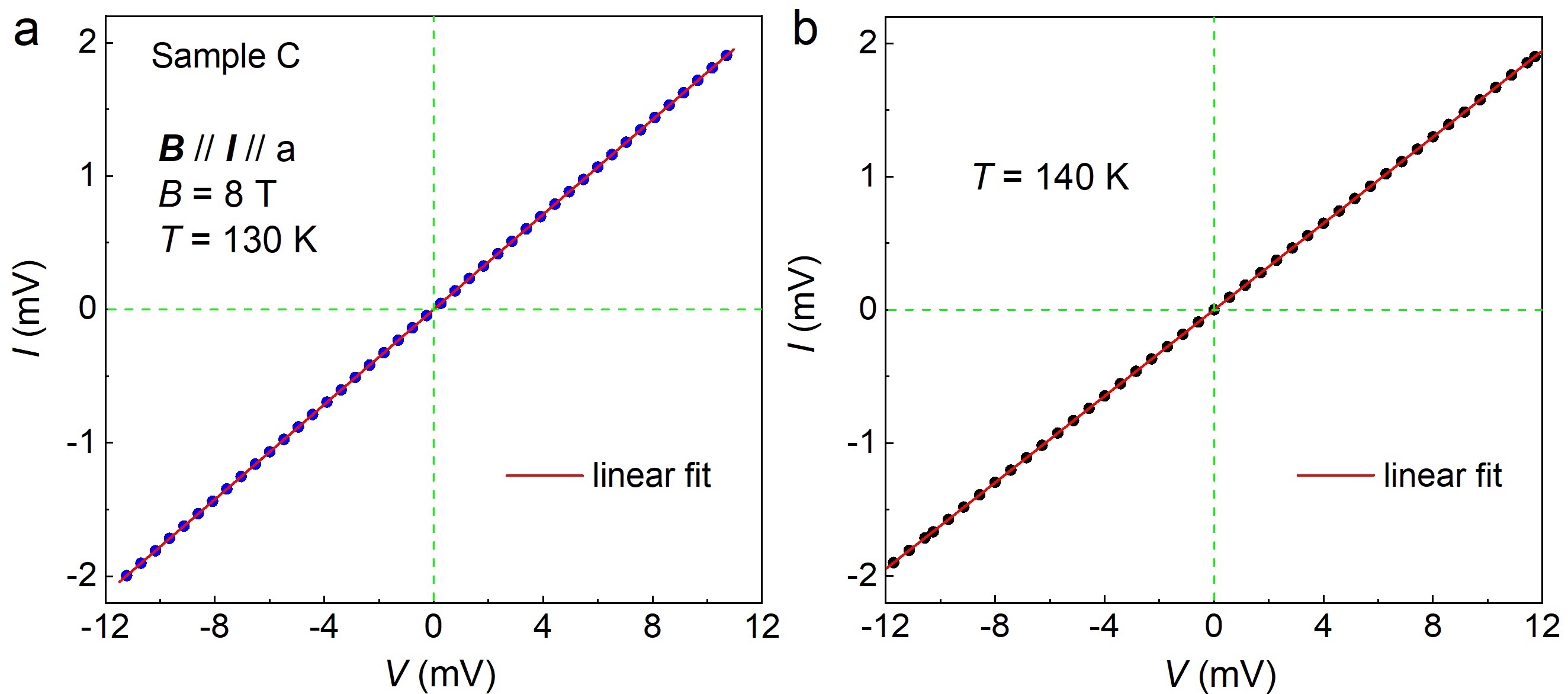}
	\caption{$I$-$V$ curves of sample C measured in 8 T at 130 K (a) and 140 K (b) with $B \parallel a$. Solid lines are linear fits to the data. The nonlinear coefficient $\gamma$ is zero within experimental resolution.
	}
	\label{fig:HighTp-highT}
\end{figure}

A samples with $T_{\rm p}$ = 133 K is no longer a nodal-line semimetal and an ordinary ellipsoidal Fermi surface is realized \cite{Tang2019}. Hence, nonlinear $I$-$V$ curves are not expected. The plots of $\rho_{xx}(T)$ and $\rho_{xx}(B)$ are shown in Figs. \ref{fig:HighTp}a and \ref{fig:HighTp}b for such a sample (sample C). The measured $I$-$V$ curves are completely linear within our resolution at 2 K in any magnetic field, as shown in Figs. \ref{fig:HighTp}c and \ref{fig:HighTp}d. 
Also, as depicted in Fig. \ref{fig:HighTp-highT}, no nonlinearity in $I$-$V$ curves was observed near $T_p$, at which the chemical potential is supposed be close to the Dirac point and the Hall coefficient shows a sign change \cite{Tang2019}. Note that the $\rho_{xx}$ value at $T_p$ in sample C is roughly an order of magnitude smaller than that of $T_p$ = 0 K samples, which is probably due to the thermal activation of carriers across the Dirac point with the thermal energy of the order of 10 meV.

\subsection{Dismissing trivial effects that could cause nonlinear transport}

For the longitudinal transport configuration, the current jetting effect \cite{Liang2018a} often comes into discussion. However, even though the current jetting can cause spurious negative MR, it does not a priori lead to a nonlinearity. 
Indeed, our multi-terminal experiment, in which a homogeneous current flow was ensured with the current contacts covering the whole width of a sample, found no difference as described below. 
A non-Ohmic contact resistance can be excluded as the origin of the pronounced nonlinearity in longitudinal magnetic fields, because the contact resistance is not expected to depend strongly on the magnitude and orientation of the magnetic field. Joule heating can indeed give rise to additional nonlinearity for currents above 2 mA in our typical samples, so we performed our experiments well below this threshold current.

\subsection{Discussion on the current jetting effect}

\vspace{-3mm}
\begin{figure}[h]
	\centering
\includegraphics[width=17 cm]{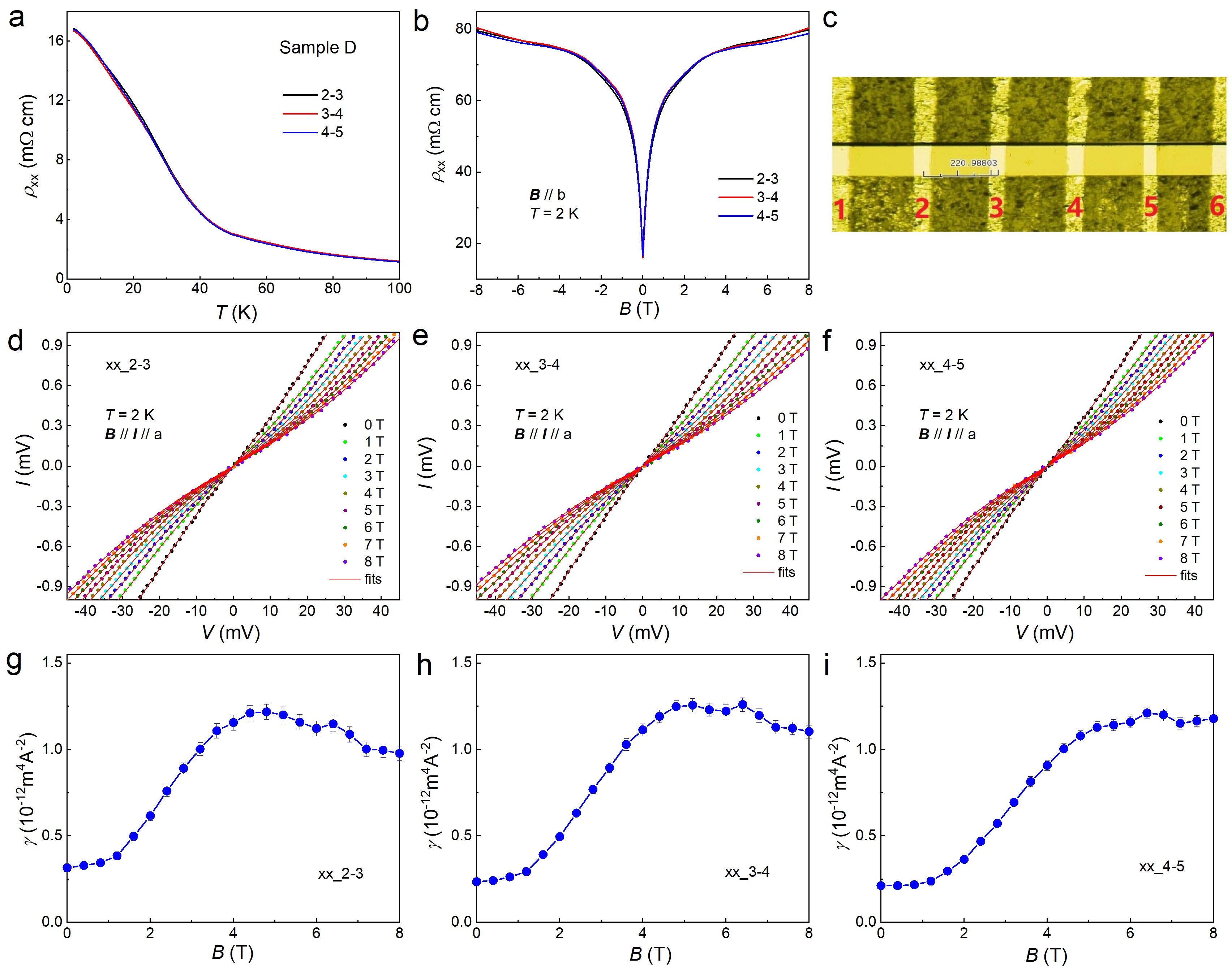}
	\caption{\textbf{Ruling out current jetting as the cause of the nonlinear transport.} (a, b) Temperature and magnetic field dependencies of the resistivity for three different pairs of terminals made on sample D. (c) A photograph of sample D to show the numbering of electrodes; the scale in the picture is an example of the optical microscope measurement of the distance between the center of electrodes in $\mu$m. (d-i) Magnetic-field dependence of the $\gamma$ values obtained from the fits of the $I$-$V$ curves. Error bars are standard error of the regression.
	}
	\label{fig:Jetting}
\end{figure}

In an Ohmic system with an arbitrary shape, the local voltage as determined by Laplace’s equation is always proportional to the applied current, even when the current distribution is inhomogeneous~\cite{Levin1997}. This means that an inhomogeneity of the current distribution itself will not cause nonlinearity. 
Hence, a global inhomogeneity in the current distribution, which might be generated by current jetting effect, would not explain the nonlinear $I$-$V$ curves observed in ZrT$_5$, as long as inhomogeneous Joule heating does not affect the $I$-$V$ curves. Here, we have performed experiments to directly confirm that the current jetting effect cannot explain the observed nonlinear transport.

Current jetting refers to the focusing of the current density into a narrow beam parallel to the magnetic field direction arising from the field-induced anisotropy of the conductivity \cite{Liang2018}. The current jetting effect usually happens in high mobility samples with point-shape contacts; when the conductivity anisotropy increases with the magnetic field, the current beam becomes narrower, causing the resistivity probed at the sides of a sample to become smaller in higher magnetic fields; this results in an apparent negative magnetoresistance (MR) \cite{Arnold2016}. To avoid current jetting, we prepared a sample [the $\rho_{xx}(T)$ and $\rho_{xx}(B)$ curves are shown in Figs. \ref{fig:Jetting}a-b] in which the electrodes are made to cover the whole transverse surface of the samples, as pictured in Fig. \ref{fig:Jetting}c, and multi-terminal measurements were performed to check the homogeneity of the current flow. As one can see in Figs. \ref{fig:Jetting}a and \ref{fig:Jetting}b, the transport results obtained between different pairs of terminals agree very well, confirming the homogeneity of the current flow through the sample. Also, the $\rho_{xx}(B)$ data shown in Fig. \ref{fig:Jetting}b is essentially symmetric with respect to $B$, giving confidence in a homogeneous contact resistance across the width (since an inhomogeneous contact usually results in an adverse Hall voltage mixing). The nonlinear $I$-$V$ curves and the peculiar $\gamma(B)$ behavior were reproduced in all parts of the samples, giving evidence that the magnetic-field-induced enhancement of $\gamma$ is not due to current jetting.

\subsection{Reproducibility of the strong angular dependence}

\begin{figure}[h]
	\centering
\includegraphics[width=14 cm]{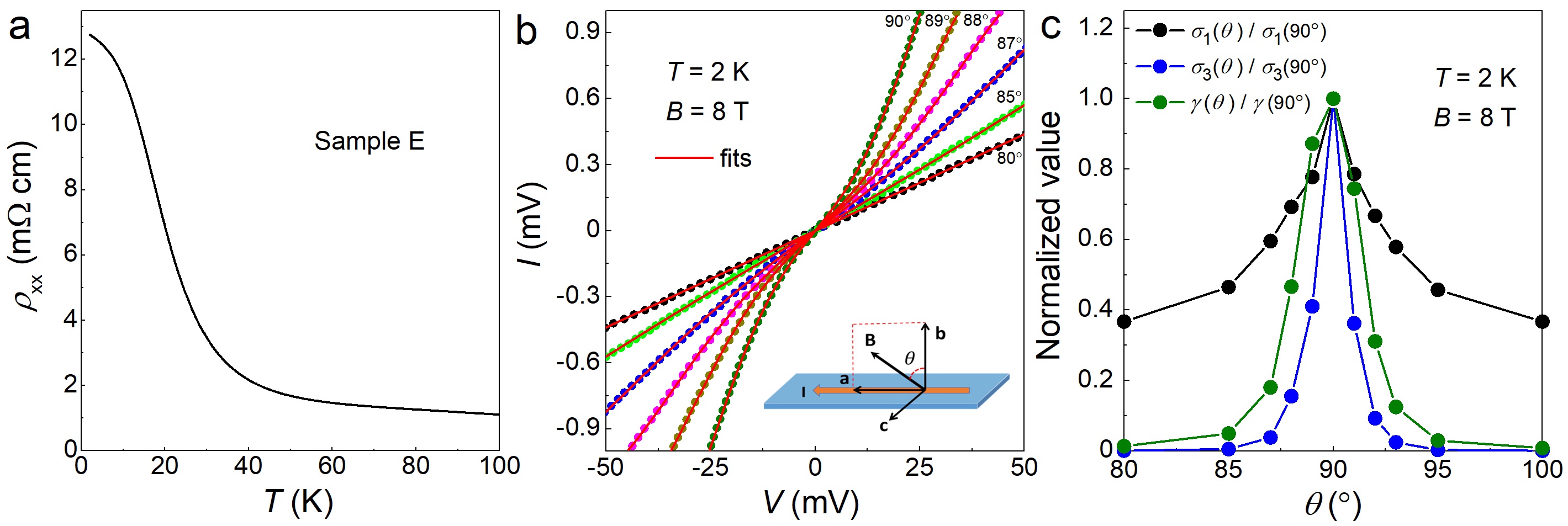}
	\caption{\textbf{Nonlinear $I$-$V$ curves in sample E.} (a) $\rho(T)$ behavior of sample E. (b) $I$-$V$ curves measured under the magnetic field rotated in the $ab$ plane, measured at 8 T and 2 K; inset depicts the measurement configuration. (c) Dependence of  $\sigma_1$, $\sigma_3$, and $\gamma$ ($\equiv \sigma_3 / \sigma_1^3$) on the magnetic-field angle $\theta$ changed in the $ab$ plane. The vertical axis is normalized with the maximum values, which are: $\sigma_1^{\rm max}$ = 29.4 $\Omega^{-1}$cm$^{-1}$, $\sigma_3^{\rm max}$ = 40.8 $\Omega^{-3}$A$^{-2}$cm, and $\gamma^{\rm max}$ = 16.6$\times 10^{-12}$ m$^{4}$A$^{-2}$.
	}
	\label{fig:sampleE}
\end{figure}

The strong angular dependence of the nonlinearity in the $I$-$V$ curves were reproduced in another sample, sample E, as shown in Fig. \ref{fig:sampleE}.

\subsection{Nonlinear $I$-$V$ characteristics in 0 T observed in all the samples}

\begin{figure}[h]
	\centering
\includegraphics[width=16.5 cm]{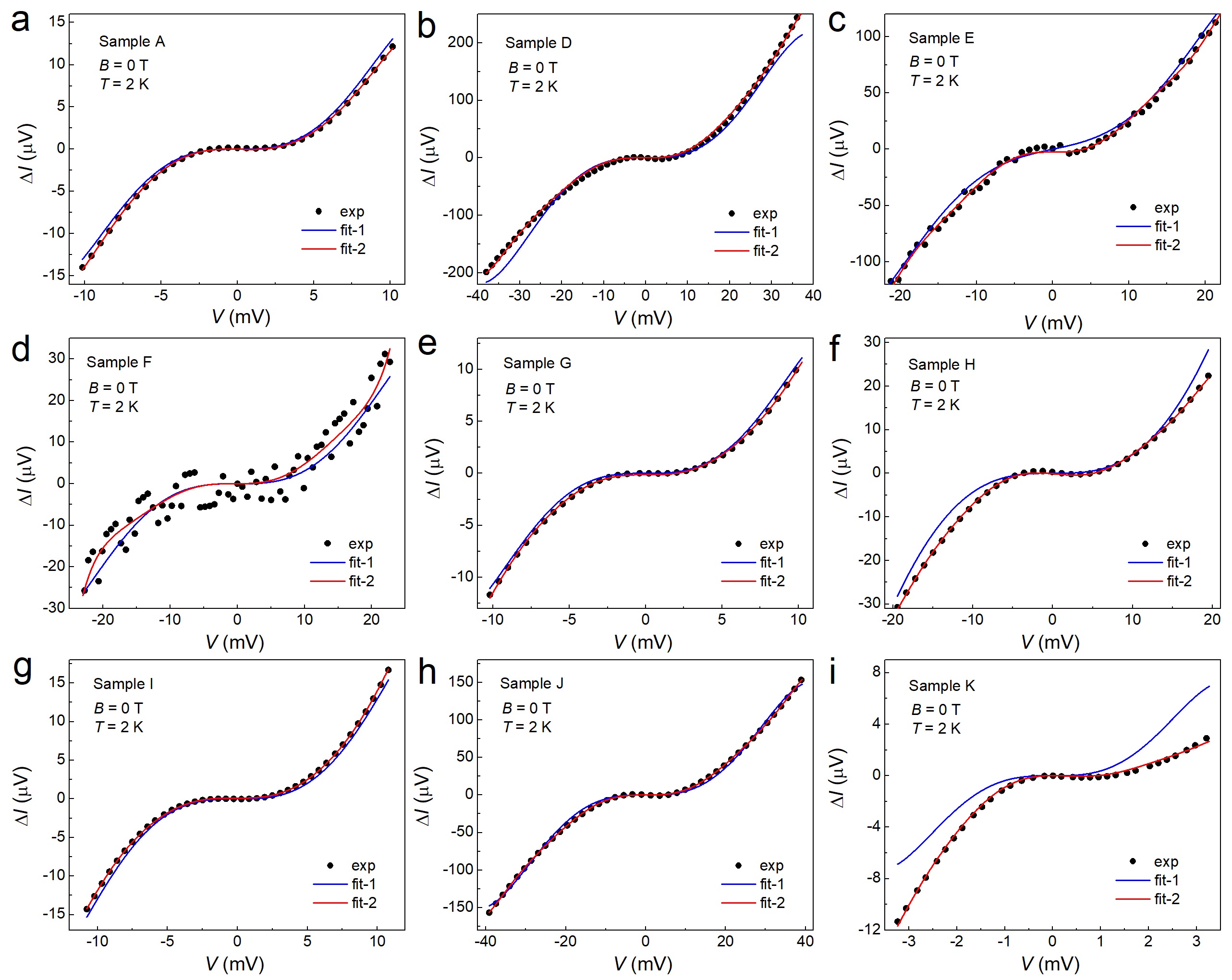}
	\caption{\textbf{Nonlinearity in the $I$-$V$ curves at 0 T.} (a-i) Deviation of the $I$-$V$ characteristics from the linear behaviour measured at 2 K without magnetic fields for all 9 samples. The $I(V)$ behavior is fitted with $I = a V + b_3 V^3 + b_5 V^5$ and $\Delta I \equiv I - aV$ is shown. The calculated $\Delta I = b_3 V^3 + b_5 V^5$ is shown with blue lines (fit-1). The deviation is also fitted with $\Delta I = c_2 V^2+ c_3 V^3 + c_5 V^5 + c_7 V^7 + c_9 V^9$ and the result is shown with red lines (fit-2). The $V^2$ term in fit-2 is to take care of the small nonreciprocity due to imperfect contacts; since the nonlinearity in 0 T is much smaller than that in high magnetic fields, a good fitting of the 0-T data needs this nonreciprocal term as well as higher odd-order terms. The $\gamma$ values at 0 T shown in Fig. 3 of the main text are obtained from fit-1 to keep consistency with the nonlinear analysis of the higher magnetic field data.
	}
	\label{fig:0T}
\end{figure}

To demonstrate the presence of nonlinearity in 0 T, we show in Fig. \ref{fig:0T} the plots of $\Delta I$ vs $V$, where $\Delta I$ is obtained by subtracting the linear term (which is the first term of the polynomial fit) from the measured $I$ for a given $V$. Even though the magnitude of the nonlinearity varies by a factor of $\sim$10, the deviation from linearity is visible in all the samples. The $\gamma$ values to represent this nonlinearity are plotted in Fig. 3 of the main text.

\subsection{Effect of magnetic fields on hot-carrier transport}

As mentioned in the main text, hot-carrier transport can be responsible for the nonlinear transport in 0 T due to the electric-field concentration between charge puddles. In this context, it is useful to mention that the effect of magnetic fields on hot-carrier transport has been studied for plain semiconductors. When a strong magnetic field is applied parallel to the electric field, the heating of the electron gas is eliminated \cite{Zlobin1972, Pinchuk1980}.
%[Pinchuk, I.I. "Hot Electrons in Semiconductors in Longitudinal Quantizing Magnetic Fields." Phys. Status Solidi B 97 (1980) 355; Zlobin, A.M., and Zyryanov, P.S. "Hot electrons in semiconductors subjected to quantizing magnetic fields." Sov. Phys. Usp. 14 (1972) 379]. 
Hence, the nonlinearity due to hot electrons would be suppressed by a strong longitudinal magnetic field, which is opposite to what is observed in ZrTe$_5$. This supports our conclusion that the magnetic-field-induced enhancement of $\gamma$ is due to an additional mechanism, the flat-band transport.

%\clearpage

\section{Theoretical description}

	\subsection{Flat bands: magnetic field in $a$ direction} \label{Sec:Flatbands}
	In this section, we consider a magnetic field parallel to the torus plane, and show
	that there emerge extremely non-dispersing bands along the momenta parallel to the magnetic field direction.
	
	The low-energy Hamiltonian for ZrTe$_5$ at zero magnetic field is constructed in the basis states of $( |\Psi^{\uparrow}_+\rangle, |\Psi^{\uparrow}_-\rangle, |\Psi^{\downarrow}_+\rangle, |\Psi^{\downarrow}_-\rangle)$ with parity $\pm$ and spin $\uparrow/\downarrow$ as~\cite{Wang2022}
	\begin{equation}
	\begin{split}
	H(\vec{k})&=m \mathbb{1} \otimes \tau_z+\hbar (v_a k_a \sigma_z \otimes \tau_x +v_b k_b \sigma_x \otimes \tau_x + v_c k_c \mathbb{1} \otimes \tau_y)+\Delta \mathbb{1} \otimes \tau_x  -\mu \mathbb{1}. \label{Hamiltonian2}
	\end{split}
	\end{equation}
	Here the Pauli matrices $\sigma_{\alpha}$ and $\tau_{\beta}$ act on the spin and parity space, respectively, and $\Delta$ is the parameter to describe the $ab$-mirror symmetry breaking.
Upon slight doping ($m<\mu <  \Delta$), one obtains a torus Fermi surface lying in the $ab$ plane (cf. Fig.~4(a) in the main text). Since the mass gap $m$ is approximately zero at temperatures close to $T_p \sim 0$ K~\cite{Xu2018}, the mass gap is neglected. 
	
	We next apply a magnetic field parallel to the torus plane (here $\vec{B} \parallel \vec{\hat{a}}$). 
	Under the Landau gauge $ \vec{A} = (0, 0, B y)$, the Hamiltonian is written as
	\begin{align} 
	H (\vec{k}) =
	\hbar \left ( v_a k_a \sigma_z \otimes \tau_x + 
	v_b (k_b + e B y)  \sigma_x \otimes \tau_x + v_c k_c  \mathbb{1} \otimes \tau_y \right ) + \Delta  \mathbb{1} \otimes \tau_x - g B \sigma_x  \otimes  \mathbb{1}  - \mu  \mathbb{1},
	\end{align}
	with electron charge $-e$, $g \equiv g_B \mu_B/2$ with $g_B$ the g-factor and $\mu_B$ the Bohr magneton. 
	Performing the unitary transformation $U = e^{i \pi \sigma_y \otimes \mathbb{1}/4}$ ($\sigma_z \rightarrow \sigma_x$, $\sigma_x \rightarrow - \sigma_z$)
	%, we obtain 
	%\begin{align} \label{Hamiltonian_adirection}
	%&U^{\dagger} H_a (\vec{k} ) U 
	%& = \begin{pmatrix}
	%- \mu  + g B & - \hbar \left( i k_c v_c  - v_b (k_b  + B e y ) \right) + \Delta & 0 & k_a v\\
	%i k_c v  - v_b (k_b  + Be y ) + \Delta & - \mu  + g B&  k_a v  & 0 \\ 
	%0&  k_a v & - \mu - g B & - i k_c v  + v_b (k_b  + B e y ) + \Delta \\
	% k_a v & 0 &   i k_c v  + v_b (k_b  + B e y ) + \Delta &  - \mu - g B  \\ 
	%\end{pmatrix}.
	%\end{align}
	and defining a lowering operator
	as $a \equiv  \ell_B   \left (v_b (k_b +  B e y )  +  i k_c v_c \right) / \sqrt{2 v_c v_b}$ such that $[a, a^{\dagger}] = 1$, we obtain
	\begin{align} \label{Hamiltonian_adirection}
	&U^{\dagger} H (\vec{k} ) U = \begin{pmatrix}
	- \mu  + g B & - \hbar \omega_c a + \Delta & 0 & \hbar k_a v_a\\
	- \hbar \omega_c a^{\dagger} + \Delta & - \mu  + g B & \hbar  k_a v_a  & 0 \\ 
	0&  \hbar k_a v_a & - \mu - g B&\hbar \omega_c a^{\dagger}  + \Delta \\
	\hbar k_a v_a & 0 &  \hbar \omega_c a  + \Delta &  - \mu - g B \\
	\end{pmatrix}.
	\end{align}
	Here $\ell_B= \sqrt{\hbar/eB}$ is the magnetic length and $\omega_c \equiv \sqrt{2 e B v_c v_b / \hbar}$ is the cyclotron energy. Eigenstates of the Hamiltonian \eqref{Hamiltonian_adirection} are labeled by $|\psi_{n, \sigma, \tau} (k_a, k_b) \rangle$ with quantum number $n = 0, 1, \cdots $, satisfying $a^{\dagger} a |\psi_{n, \sigma, \tau} (k_a, k_b) \rangle = n|\psi_{n, \sigma, \tau} (k_a, k_b) \rangle$. 
	
With vanishing $\Delta$, the Hamiltonian is block-diagonal in the basis states of $|\psi_{n = 0}  \rangle = ( |\psi_{0, \uparrow, -} \rangle,|\psi_{0, \downarrow, +} \rangle )^T$  for $n = 0$ and $|\psi_{n}  \rangle = (|\psi_{n-1, \uparrow, +} \rangle, |\psi_{n, \uparrow, -} \rangle, |\psi_{n, \downarrow, +}\rangle, |\psi_{n-1, \downarrow, -}) \rangle )^T$ for $n \geq 1$. Therefore, the Hamiltonian is readily diagonalized within given sector $n$. 
Upon turning on $\Delta$, $|\psi_n  \rangle$ starts to couple with the neighboring states 
	$|\psi_{n-1}  \rangle$ and $|\psi_{n+1}  \rangle$. To obtain the band structure, we construct a $(4 \times M + 2 ) \times (4 \times M + 2)$ Hamiltonian matrix with the matrix size cut-off determined by integer $M$. 
Until the Landau level energies converge, we increase the size of the Hamiltonian. Note that as $B$ becomes smaller, it requires a larger size of the Hamiltonian to converge.

	To figure out the origin of the flat bands, we first note from \eqref{Hamiltonian_adirection} that when $k_a = 0$ (a sweet spot), the Hamiltonian is divided into two block diagonal $2 \times 2$ matrices $H_{\uparrow/\downarrow}$ with different spins $\sigma$ = $\uparrow$, $\downarrow$. Each of the block diagonal Hamiltonians describes a 2D Dirac point. Neglecting the Zeeman term, $H_{\uparrow/\downarrow}$ is written as
	\begin{align}
	H_{\uparrow} = 
	\begin{pmatrix}
	0   & -  \hbar \omega_c  a_{\uparrow} \\
	- \hbar \omega_c a_{\uparrow}^{\dagger} & 0 \\
	\end{pmatrix},\,\,\,\,\,\,\,\,\,\,\,
	H_{\downarrow} = 
	\begin{pmatrix}
	0   &  \hbar \omega_c  a^{\dagger}_{\downarrow} \\
	\hbar \omega_c  a_{\downarrow} & 0 \\
	\end{pmatrix}
	\end{align}
	with lowering operators $a_{\uparrow / \downarrow} \equiv \ell_{B} \left (i k_c v_c + v_b \left(\left (k_b  \mp 
	\delta k_b /2 \right) + B e y \right ) \right)/ \sqrt{2 v_c v_b}$, where $\delta k_b $ is the momentum distance of the two 2D Dirac points. $\delta k_b (k_a)\sim 2 \sqrt{ (\Delta/(\hbar v_b))^2-k_a^2}$  and $\delta k_b = 2 \Delta / (\hbar v_b)$ at $k_a = 0$. 
	Each of the block diagonal Hamiltonian can be separately diagonalized, resulting in the Landau levels. 
	Importantly, the lowest Landau levels are located at zero energy with the eigenstates
	\begin{align}
	\psi_{0, \uparrow} =\begin{pmatrix} 0\\ \psi_{0, \uparrow, -} (k_b; y, z)
	\end{pmatrix},\,\,\,\,\,\,\,\,\,\,
	\psi_{0, \downarrow} =\begin{pmatrix} \psi_{0, \downarrow, +} (k_b; y, z)\\ 0
	\end{pmatrix}. 
	\end{align}
	Here
	\begin{align} \label{supp:eigenstates}
	\psi_{0, \uparrow, \tau } (k_b; y, z) &= \left ( \frac{v_b}{\pi v_c \ell_B^2} \right)^{1/4} \exp \left ({-\frac{v_b}{2 \ell_B^2 v_c} \left ( y + \ell_B^2 \left ( k_b - \delta k_b/2 \right) \right)^2} \right)  e^{i k_b z}, \nonumber \\ \psi_{0, \downarrow, \tau } (k_b; y, z) &= \left ( \frac{v_b}{\pi v_c \ell_B^2} \right)^{1/4} \exp \left ({-\frac{v_b}{2 \ell_B^2 v_c} \left ( y + \ell_B^2 \left ( k_b + \delta k_b/2 \right) \right)^2} \right)  e^{i k_b z}.
	\end{align}
It can be seen from the exponentially decaying factors in \eqref{supp:eigenstates} that two eigenstates 
	$\psi_{0, \uparrow, \tau} (y, z)$ and $\psi_{0, \downarrow, \tau} (y, z)$ have the width of $W = \ell_B \sqrt{v_c/ v_b}$, and are spatially separated by $\Delta y = \ell_B^2 \delta k_b$ in $y$ direction. 
	%$\psi_{0, \up, \tau}$
	
	%given $k_b$, the two eigenstates are spatially separated by $2 \ell_B^2 \Delta / v_b$ in $y$ direction (one at $y = - \ell_B^2 k_b + \ell_B^2 \Delta /v_b $ and the other at $y =  - \ell_B^2 k_b  -\ell_B^2 \Delta /v_b$. 
	We now turn on finite $k_a$ and the Zeeman term. Then, the two lowest Landau levels start to couple and deviate from the zero energy. In the degenerate perturbation theory, we construct the effective Hamiltonian 
	% are moving to $k_a$ away from $k_a  = 0$ and two lowest Landau levels start to couple. In order to see how they are coupled, we construct a projected Hamiltonian only for the lowest Landau levels. 
	\begin{align}
	H (k_a, k_b) &= \begin{pmatrix}
	g B   &    k_a v_a \int dy \psi_{0, \uparrow, -}^{*} (k_b; y, z) \psi_{0, \downarrow, +} (k_b; y, z) \ \\ 
	k_a v_a \int dy \psi_{0, \downarrow, +}^{*} (k_b; y, z) \psi_{0, \uparrow, -} (k_b; y, z)   & - g B \\
	\end{pmatrix}  \\ 
	&\sim \begin{pmatrix}
	g B   &   2 k_a v_a e^{-  (\ell_B \delta k_b)^2 v_b / (2 v_c)} \ \\ 
	2 k_a v_a e^{-  (\ell_B \delta k_b)^2 v_b / (2 v_c)}   & -g B \\
	\end{pmatrix}.  \label{supple:effectiveHam}
	\end{align}
	We have assumed that the two eigenstates are well-separated such that $W \ll \Delta y$ ($\ell_B \gg 1/\delta k_b$ and hence $B \ll \hbar \delta k_b^2 v_b/v_c$).
	 Under this assumption, the overlap between the eigenstates vanishes exponentially as 
	$e^{-  (\ell_B \delta k_b)^2 v_b / (2 v_c)}$. In the absence of the Zeeman term, 
	the energies of the lowest Landau levels becomes $\epsilon_{k_a}^{\pm} \sim \pm 2 k_a v_a e^{-  (\ell_B \delta k_b)^2 v_b / (2 v_c)}$ and hence the velocities of the flat bands behaves as $v_f^{\pm} \sim \pm 2 v_a e^{-  (\ell_B \delta k_b)^2 v_b / (2 v_c)}\left ( 1 +2 (k_a \ell_B)^2 \frac{v_b}{v_c}\right)$. Using the parameters $\Delta \simeq 19.1$ meV, $v_a \simeq 6.9 \times 10^5 $ m/s, $v_b \simeq 0.43 \times 10^5$ m/s, $v_c \simeq 1.7 \times 10^5 $ m/s obtained from the quantum oscillation measurements~\cite{Wang2022}, the velocity of the lowest Landau-level bands is exponentially suppressed in all magnetic field $B \ll \hbar \delta k_b^2 v_b/(v_c e) \sim 80$ T where our measurements are done ($B < 8$ T). 
	
 The Landau-level bands in the absence of the Zeeman term are plotted in Fig.~4c of the main text. 
This calculation has been done using the parameters $B = 1$ T,  $\Delta= 19.1$ meV, and $v_a = 6.9 \times 10^5$ m/s. The two lowest Landau levels are degenerate with the exponentially small splitting. 

We next consider the effect of the Zeeman term. The diagonalization of \eqref{supple:effectiveHam} leads to the energies $\epsilon_f^{\pm}$ and the velocities $v_f^{\pm}$ of the lowest Landau levels as
\begin{align} \label{flatbandZeeman}
\epsilon_f^{\pm} &=\pm \sqrt{(g B)^2 + (2 k_a v_a)^2 e^{-  (\ell_B \delta k_b)^2 v_b / v_c}} \nonumber \\ 
v_f^{\pm} &= \pm v_a \frac{4 k_a v_a e^{-  (\ell_B \delta k_b)^2 v_b / v_c}}{\sqrt{(g B)^2 + (2 k_a v_a)^2 e^{-  (\ell_B \delta k_b)^2 v_b / v_c}}} \left ( 1 +2 (k_a \ell_B)^2 \frac{v_b}{v_c}\right). 
\end{align}
The Landau-level bands in the presence of the Zeeman term are displayed in Fig.~4d of the main text. We have used $g_B = 2$ for the plot. As seen both in the plot and the analytic expressions \eqref{flatbandZeeman}, a small $k_a$-dependent splitting occurs due to the Zeeman term.

\subsection{Transport calculations}

As the flat-band region of the lowest Landau level can host a large number of states with small velocity, it will have a large impact on the transport properties in the quantum limit. To calculate the nonlinear transport, we evaluate the Boltzmann equation up to third order in electric field. We will mainly work in the relaxation time approximation where the Boltzmann equation can be written as
\begin{align} \label{Boltzmann}
e \left( \vec{E + \vec{v}_k \times \vec{B}} \right) \frac{\partial f}{\partial \vec{k}} = \frac{f - f_0}{\tau},
\end{align}
where we have set $\hbar =1$.  The distribution function $f = f_0 + f_1 +f_2 + ...$ can be expanded in powers of the electric field with $f_m \sim E^m$ and $f_0$ being the Fermi-Dirac distribution. 

Note that the scattering time $\tau$ is not known here and it is expected to depend strongly not only on the disorder level but also on the magnetic field. Therefore it is important that we consider below only quantities which are independent of $\tau$.

\subsubsection{Boltzmann equation without magnetic field}

We first study our system in the absence of magnetic field and determine the nonlinear response when applying an electric field in the $a$-direction. We will first present the calculation for the isotropic case, meaning $v_a=v_b=v_c=v$. The results can be generalized to an anisotropic system via a rescaling of the momenta by $k_i \rightarrow k_i/v_i$. In the isotropic case the band structure is best described in cylindrical coordinates with $k_a= k_r\ \mathrm{cos}(\phi)$ and $k_b= k_r\  \mathrm{sin}(\phi)$. The dispersion of the relevant band is given by
\begin{align} \label{dispersion}
\epsilon_{\vec{k}} = \sqrt{m^2 + \hbar^2 v^2 k_c^2 + \left(\hbar v k_r - \Delta \right)^2 }.
\end{align}
As the mass is fine-tuned in the experiments to $m \approx 0$, we will neglect the mass term in the following, which allows us to evaluate the Boltzmann equation analytically. Numerical solutions show that the contributions from the mass term are small and can be neglected for $|m| \ll \mu$. Thus, a nodal line is located at $k_r=\Delta/(\hbar v)$ and $k_c=0$.\\
Using the relaxation time approximation, the first-order distribution function at zero temperature (setting $\hbar = 1$ for now) can be written as
\begin{align} \label{ZeroField1stOrder}
f_1 = e \tau E  \frac{\partial \epsilon_{\vec{k}}}{\partial k_a} \frac{\partial f_0}{\partial \epsilon_{\vec{k}}} = e \tau E \left( \frac{v (v k_r - \Delta)\mathrm{cos}(\phi)}{\sqrt{m^2 + v^2 k_c^2 + \left(\Delta - v k_r \right)^2 }} \right) \delta(\mu - \epsilon_{\vec{k}}).
\end{align}
From this, we obtain the first-order current in the $a$-direction
\begin{align} \label{ZeroField1stOrderCurrent}
j^{(a)}_1 = -e \int \frac{\mathrm{d}^3\vec{k}}{(2\pi)^3} v_a f_1.
\end{align}
As we are interested in the longitudinal conductivity, there is no contribution arising from anomalous velocities due to the Berry curvature, because these terms are of the form $\vec{\Omega} \times \vec{E}$ and they can only generate a current perpendicular to the applied field. Putting \eqref{ZeroField1stOrder} into \eqref{ZeroField1stOrderCurrent}, we obtain
\begin{align} \label{ZeroField1stOrderCurrentResult}
j^{(a)}_1 = -e \int \frac{\ \mathrm{d}k_r \mathrm{d}k_c \mathrm{d}\phi}{(2\pi)^3} \left(\frac{\partial \epsilon_{\vec{k}}}{\partial k_a}\right)^2 \delta(\mu - \epsilon_{\vec{k}}) = \frac{e^2 \tau E}{8 \pi v} \mu \Delta.
\end{align}
Restoring $\hbar$ and the anisotropic velocities (by rescaling the integration measure), we obtain the longitudinal conductivity as
\begin{align} \label{ZeroField1stOrderConductivity}
\sigma^{(aa)}_1 = \frac{e^2\tau \mu \Delta v_a}{8 \pi \hbar^3 v_b v_c}.
\end{align}
Continuing with the expansion in powers of $E$, the third-order current is given by integrating over the third derivative of the dispersion
\begin{align} \label{ZeroField3rdOrderCurrentResult}
j^{(a)}_3 = -e \int \frac{\ \mathrm{d}k_r \mathrm{d}k_c \mathrm{d}\phi}{(2\pi)^3} \frac{\partial \epsilon_{\vec{k}}}{\partial k_a} \frac{\partial^3 \epsilon_{\vec{k}}}{\partial k_a^3} \delta(\mu - \epsilon_{\vec{k}}) = \frac{3 e^4 \tau^3 E^3 v_a^3 \Delta}{128 \pi \hbar^3 v_b v_c \mu} \left( 5- \frac{8 \Delta}{\sqrt{\Delta^2 - \mu^2}} \right) = \sigma^{(aa)}_3 E^3.
\end{align}
As stated before, the scattering time $\tau$ is unknown. Thus, we characterize the size of the nonlinearity by a quantity independent of the scattering rate 
\begin{equation}
\gamma_3 = \frac{\sigma^{(aa)}_3}{\left( \sigma^{(aa)}_1\right)^3} = \frac{12 \pi^2 \hbar^6 v_b^2 v_c^2}{e^2 \Delta^2 \mu^4}\left( 5- \frac{8 \Delta}{\sqrt{\Delta^2 - \mu^2}} \right).
\label{Gamma3Definition}
\end{equation}
By integrating over the torus Fermi surface, we can obtain the charge carrier density as a function of chemical potential
\begin{equation}
n(\mu) = \frac{\Delta \mu^2}{4 \pi \hbar^3 v_a v_b v_c}.
\label{DensityZeroField}
\end{equation}
Using this, we can write the nonlinear coefficient as a function of density at zero magnetic field
\begin{equation}
\gamma_3(B=0) = \frac{3}{4 e^2 v_a^2} \left( \dfrac{3}{n^2} +16 \pi \frac{\hbar^3 v_a v_b v_c}{\Delta^3 n}\right).
\label{Gamma3ZeroField}
\end{equation}
Repeating the same calculation for a Dirac semimetal with dispersion $\epsilon_{\vec{k}}=\sqrt{v_a k_a+v_b k_b+v_c k_c}$ yields a similar relation $\gamma_3(B=0) =\frac{6}{5 e^2 v_a^2 n^2}$.

\subsubsection{Boltzmann equation for Landau levels}

\NEW{Now we apply this formalism to the nodal line in magnetic fields. Instead of using the three-dimensional Boltzmann equation \eqref{Boltzmann}, which is only valid for weak magnetic fields,  we have to use for strong magnetic fields the Boltzmann equation directly for the band structure obtained in a strong magnetic field, using the 
 dispersion $\epsilon_{n,k}$ of the Landau levels discussed in Sec.~\ref{Sec:Flatbands}, see also Fig.~4d in the main text. For electric fields parallel to the magnetic field and within relaxation time approximation, we can simply use for each band
\begin{align} \label{BoltzmannLL}
e  E \frac{\partial f_{n,k}}{\partial k} = \frac{f_{n,k} - f_0(\epsilon_{n,k}-\mu)}{\tau},
\end{align} 
where $f_0$ is the equilibrium Fermi function and the three-dimensional current density parallel to $B$ is computed from 
$j_\|=\frac{e}{2 \pi l_B^2} \sum_n \int \frac{dk}{2 \pi}  f_{n,k} \partial_k \epsilon_{n,k}$ where the prefactor keeps track of the degeneracy of Landau levels with $l_B=\sqrt{\hbar/(e B)}$. Importantly, $\mu$ has to be recomputed from the charge density $\frac{1}{2 \pi l_B^2} \sum_n \int \frac{dk}{2 \pi}  f_0(\epsilon_{n,k}-\mu)$  for each $B$ field. The parameter $\tau$ is expected to depend strongly on the magnetic field and the electron density, but we will use \eqref{BoltzmannLL} only to compute $\gamma_3(B)$ which, by construction, is independent of $\tau$ within the relaxation time approximation. An equation equivalent to \eqref{BoltzmannLL} (for a model with Weyl nodes) has recently been derived by Wang {\it et al.} \cite{Wang2022} starting from a Lindblad-version of the relaxation time approximation by considering the limit of large magnetic fields for electric fields of arbitrary strength. Ref.~\cite{Wang2022} also discusses how \eqref{BoltzmannLL} can be solved analytically for arbitrarily large electric field but we will only need  perturbative solutions below.
}

Within this approximation, the 1st-order distribution function for the $n$th band for an electric field applied in the $a$-direction is given by
\begin{align} \label{Boltzmann1stOrder}
f_{1,n} = e \tau E \frac{\partial \epsilon_{n,k_a}}{\partial k_a} \frac{\partial f_{0,n}}{\partial \epsilon_{n,k_a}}.
\end{align}
Summing up all these contributions and multiplying them with the corresponding velocity gives the 1st-order current in the $a$-direction
\begin{align} \label{1stOrderCurrent}
j_1 = -\frac{e^2 \tau E}{2 \pi l_B^2} \sum_n \int \frac{dk_a}{2\pi} \left( \frac{\partial \epsilon_{n,k_a}}{\partial k_a}\right)^2 \frac{\partial f_{0,n}}{\partial \epsilon_{n,k_a}} = \sigma_1 E,
\end{align}
with the magnetic length $l_B=\sqrt{\hbar/(e B)}$. Note that additional dependence on $B$ enters this formula as the Landau level dispersion $\epsilon_{n,k_a}$ as well as the chemical potential $\mu$ will depend on the magnetic field. Since the velocity $ \partial \epsilon_{n,k_a} / \partial k_a$ is small in the flat-band limit, this current will be drastically reduced when this limit is reached, assuming a constant $\tau$.\\
Extending the relaxation time approximation to third order, we obtain  
\begin{align} \label{Boltzmann3rdOrder}
f_{3,n} = e^3 \tau^3 E^3 \frac{\partial^3 f_{0,n}}{\partial k_a^3} = e^3 \tau^3 E^3 \left( \frac{\partial^3 \epsilon_{n,k_a}}{\partial k_a^3} \frac{\partial f_{0,n}}{\partial \epsilon_{n,k_a}} + 2 \frac{\partial^2 \epsilon_{n,k_a}}{\partial k_a^2} \frac{\partial}{\partial k_a} \frac{\partial f_{0,n}}{\partial \epsilon_{n,k_a}} + \frac{\partial \epsilon_{n,k_a}}{\partial k_a} \frac{\partial^2}{\partial k_a^2} \frac{\partial f_{0,n}}{\partial \epsilon_{n,k_a}} \right) .
\end{align}
Using this and rewriting the momentum derivatives inside the integral, the current cubic in the electric field is given by
\begin{align} \label{3rdOrderCurrent}
j_3 = -\frac{e^4 \tau^3 E^3}{2 \pi l_B^2} \sum_n \int \frac{dk_a}{2\pi}  \frac{\partial \epsilon_{n,k_a}}{\partial k_a}  \frac{\partial^3 \epsilon_{n,k_a}}{\partial k_a^3} \frac{\partial f_{0,n}}{\partial \epsilon_{n,k_a}} = \sigma_3 E^3.
\end{align}
For all our calculations we work with a fixed density $n(\mu) =\sum_n \int \frac{dk_a}{4\pi^2 l_B^2} f(\mu-\epsilon_{n,k_a})$, meaning that the chemical potential will depend on the magnetic field $\mu=\mu(B)$. As the magnetic field is increased, the number of states increases for all bands because of the degeneracy of the Landau levels. This results in a decrease in the chemical potential and allows us to study the properties of the lowest Landau level.\\
Again, we define the nonlinear coefficient by the scattering-rate-independent quantity $\gamma_3$ given by
\begin{equation}
\gamma_3(B) = \frac{\sigma_3}{\left( \sigma_1 \right)^3} = \frac{(2 \pi \hbar)^4}{e^4 B^2} \frac{\sum_n \int \mathrm{d}k_a \frac{\partial \epsilon_{n,k_a}}{\partial k_a} \frac{\partial^3 \epsilon_{n,k_a}}{\partial k_a^3}\frac{\partial f_0}{\partial \epsilon_{n,k_a}} }{\left(\sum_n \int \mathrm{d}k_a \left( \frac{\partial \epsilon_{n,k_a}}{\partial k_a} \right)^2 \frac{\partial f_0}{\partial \epsilon_{n,k_a}} \right)^3} \,,
\label{Gamma3LandauLevels}
\end{equation}
which is used for the calculations of the curves shown in Fig. 5 of the main text.

\subsection{Dependence on the magnetic-field angle} \label{sec:field_angle}

%In this section, we consider the dependencies of the band structure and the $\gamma$ value on the magnetic-field angle changed in the torus plane ($ab$ plane). 

\NEW{In this section, we consider the magnetic-field-direction dependence of the flat bands when the magnetic field is rotated both in the torus plane ($ab$ plane) and in the $ac$ plane (which is orthogonal to the torus plane). Through direct calculations of the band structure, we show that flat bands acquire a velocity (i.e. slope) in a drastic manner when the in-plane magnetic field deviates away from $\vec{B} \parallel \hat{a}$. This originates from the strong anisotropy in the zero-field Fermi velocity ($v_a = 16 v_b$) in the torus plane. In contrast, such an effect is relatively weak when the magnetic field is tilted away in the out-of-plane direction ($ac$ plane) due to the smaller velocity anisotropy ($v_a = 4 v_c$).}

The magnetic field in the $ab$ plane can be parameterized with angle $\theta$ as $\vec{B} = B (\sin \theta, 0, \cos \theta)$. 
We start with \eqref{Hamiltonian2} reparameterizing in terms of momentum $k_{\parallel} = k_b \cos \theta + k_a \sin \theta$ and $k_{\perp} = - k_b \sin \theta + k_a \cos \theta$, parallel and perpendicular to the magnetic field, respectively. 
Under the Landau gauge $\vec{A} = B (0, r_{\perp}, 0)$ with $r_{\perp} = x \cos \theta - z \sin \theta$ representing the real space coordinate  perpendicular to the magnetic field, the Hamiltonian becomes
\begin{equation}
	\begin{split}
	H(\vec{k})&=\hbar \left[ v_a (k_{\parallel} \sin \theta +k_{\perp} \cos \theta ) \sigma_z \otimes \tau_x +
	v_b  (k_{\parallel} \cos \theta - k_{\perp} \sin \theta ) \sigma_x \otimes \tau_x + v_c ( k_c + e B r_{\perp} ) \mathbb{1} \otimes \tau_y \right]
	 \nonumber \\ & 
	+\Delta \mathbb{1} \otimes \tau_x - g \vec{B}\cdot \vec{\sigma} \otimes  \mathbb{1}. \label{Hamiltonianangle}
	\end{split}
\end{equation}
	
\begin{figure}[h]
	\centering
\includegraphics[width=7cm]{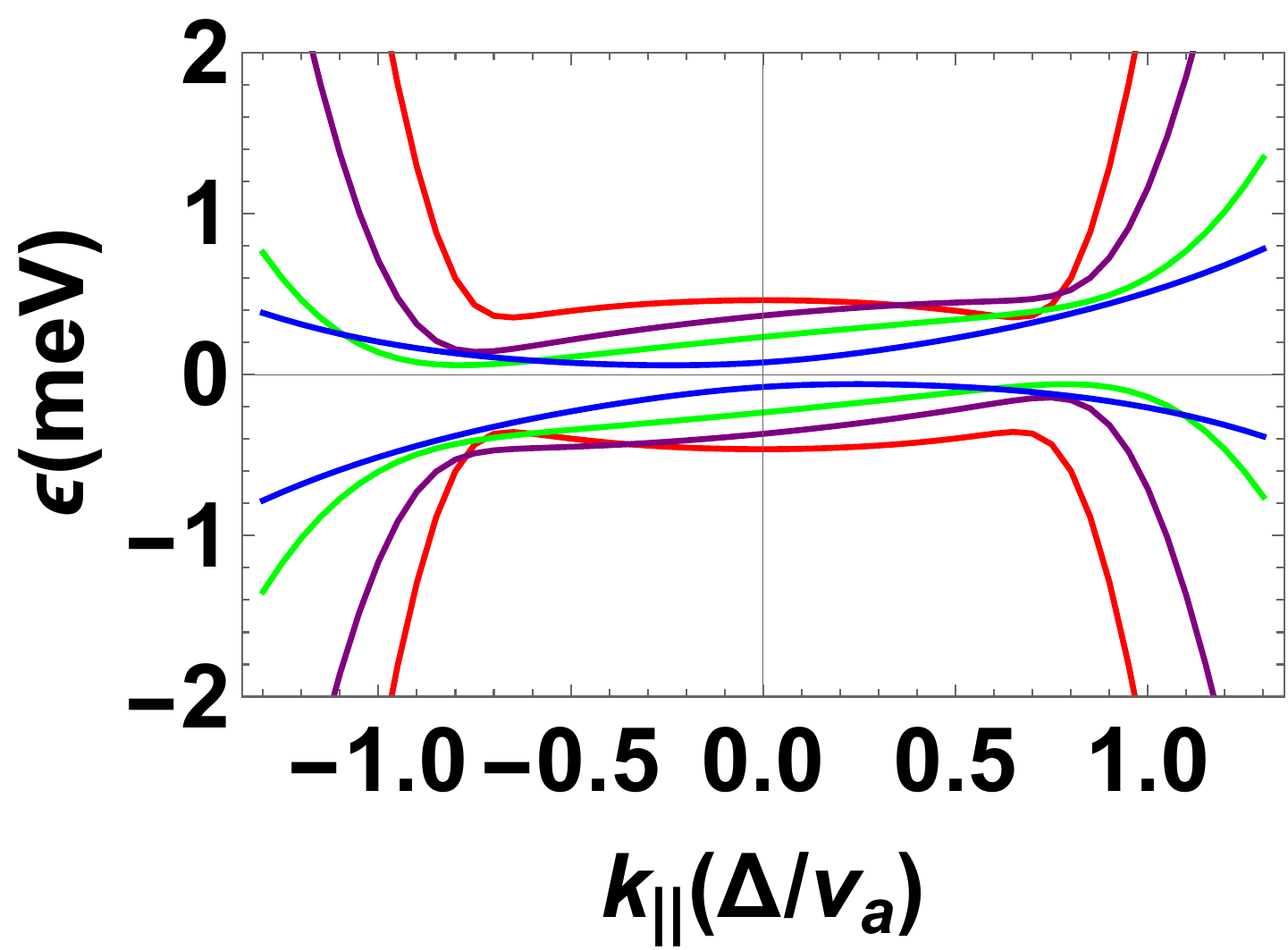}
	\caption{{\bf Dependence of the lowest Landau levels on the magnetic-field angle $\theta$ in the $ab$ plane.} The lowest Landau levels are plotted for various magnetic-field orientations  $\theta = 90, 87.5, 85, 80^{\circ}$ (red, purple, green, blue, respectively) in the $ab$ torus plane. The magnitude of the magnetic field is $8$ T. 
We have used the g factor $g_B = 2$, $\Delta= 19.1$ meV, $v_b = 0.43 \times 10^5$ m/s, $v_a = 6.9 \times 10^5$ m/s, and $v_c = 1.7 \times 10^5$ m/s.}
	\label{SuppleFig:Bandstructure}
\end{figure}

Performing the unitary transformation $U = e^{i \left (\pi/4 - \tan^{-1} (v_b \cot(\theta)/v_a )\right) \sigma_y \otimes \tau_y /2}$
and defining a lowering operator  
	as $a \equiv  \ell_B   \left (i v_c (k_c +  B e r_\perp )  -   k_\perp v_\perp - k_\parallel (v_a^2-v_b^2) \sin \theta \cos \theta/v_{\perp} \right) / \sqrt{2 v_c v_\perp}$ such that $[a, a^{\dagger}] = 1$, we obtain the Hamiltonian, 
\begin{align} \label{Hamiltonian_abdirection}
	&U^{\dagger} H (\vec{k} ) U = \begin{pmatrix}
	- g B \cos \theta_1 & - \hbar \omega_c a + \Delta & g B \sin \theta_1 & \hbar k_{\parallel} v_a v_b / v_{\perp}\\ 
	- \hbar \omega_c a^{\dagger} + \Delta & - g B \cos \theta_1 &  \hbar k_{\parallel} v_a v_b / v_{\perp}& g B \sin \theta_1 \\ 
	g B \sin \theta_1&   \hbar k_{\parallel} v_a v_b / v_{\perp} & g B \cos \theta_1 &\hbar \omega_c a^{\dagger}  + \Delta \\
 \hbar k_{\parallel} v_a v_b / v_{\perp}& g B \sin \theta_1 &  \hbar \omega_c a  + \Delta & g B \cos \theta_1 \\
	\end{pmatrix}.
	\end{align}
	Here, $\theta_1 \equiv \theta - \tan^{-1} (v_b \cot(\theta)/v_a )$, $v_{\perp} = \sqrt{v_a^2 \cos^2 \theta + v_b^2 \sin^2 \theta}$, and 
	the cyclotron energy $\omega_c = \sqrt{2 e B v_{\perp} v_c / \hbar}$. 
	The Hamiltonian can be diagonalized by constructing the Hamiltonian matrix with basis states $|\psi_{n, \sigma, \tau} (k_a, k_b) \rangle$ 
	as specified in the text below \eqref{Hamiltonian_adirection}. 
	
In Fig.~\ref{SuppleFig:Bandstructure}, the lowest Landau levels are plotted as a function of $k_{\parallel}$ with different angles $\theta = 90, 87.5, 85, 80^{\circ}$ of the magnetic field in the $ab$ plane. The velocity of the lowest Landau level bands is dramatically enhanced as the magnetic field is tilted away from the $a$ direction $\theta = 90^{\circ}$. This enhancement of the velocity is attributed to the large Fermi-velocity anisotropy in ZrTe$_5$, $v_a \simeq 16 v_b$. As discussed in Sec.~\ref{Sec:Flatbands}, the velocity of the lowest Landau level bands is exponentially small due to the exponentially vanishing coupling between the two 2D Dirac points. As the magnetic field is slightly tilted away from the $a$ direction, the effective velocity $v_{\perp} = \sqrt{v_a^2 \cos^2 \theta + v_b^2 \sin^2 \theta}$ perpendicular to the magnetic-field direction rapidly increases and hence the coupling, in turn the velocity, becomes dramatically enhanced.  This will lead to a fast decrease of the nonlinear coefficient $\gamma$ when the magnetic field is turned away from the $a$-direction, as seen in Fig. 1c.

\NEW{In contrast, the effect of the rotation angle on the velocity of flat bands is relatively weaker with the magnetic-field rotation in the $ac$ direction (out of the torus plane). This is attributed to the smaller velocity anisotropy in the $ac$ plane ($v_a \approx 4 v_c$) of ZrTe$_5$ samples. To obtain the band structures for the magnetic-field orientation rotating in the $ac$ plane, we perform a similar calculation to the rotation in the $ab$ plane. We parameterize a magnetic field rotating in the $ac$ plane as $\vec{B} = B( \cos \phi, \sin \phi, 0)$ with angle $\phi$. Using the Landau gauge $\vec{A} = B (0, 0, r_\perp )$ with the real space coordinate $r_\perp = y \cos \phi - x \sin \phi$ perpendicular to the magnetic field, and then performing the unitary transformation $U = e^{i \sigma_x \otimes \tau_z \frac{\pi}{4}} e^{i \sigma_y \otimes \mathbb{1} (\frac{\phi_1}{2} -  \frac{\pi}{4})} $, one obtains the Hamiltonian with angle $\phi$ 
   \begin{align} \label{Hamiltonian_acdirection}
	H_{ac} (\vec{k} ) = \begin{pmatrix}
	- g B \cos (\phi-\phi_1) &  \hbar \omega_c a^{\dagger}  - i \Delta \cos\phi \frac{v_c}{v_\perp} & g B \sin (\phi-\phi_1) & - (\hbar k_{\parallel} v_c + i \Delta \sin \phi) \frac{v_a}{v_{\perp}}\\ 
	 \hbar \omega_c a  + i \Delta \cos\phi \frac{v_c}{v_\perp} & - g B \cos (\phi+\phi_1) &  (-\hbar k_{\parallel} v_c + i \Delta \sin \phi) \frac{v_a}{v_{\perp}}& -g B \sin (\phi+\phi_1) \\ 
	 g B \sin (\phi-\phi_1)&   (-\hbar k_{\parallel} v_c - i \Delta \sin \phi) \frac{v_a}{v_{\perp}} &  g B \cos (\phi-\phi_1) & -\hbar \omega_c a  + i \Delta \cos\phi \frac{v_c}{v_\perp} \\
(-\hbar k_{\parallel} v_c + i \Delta \sin \phi) \frac{v_a}{v_{\perp}} &  -g B \sin (\phi + \phi_1)&  - \hbar \omega_c a^{\dagger}  - i \Delta \cos\phi \frac{v_c}{v_\perp} &  g B \cos (\phi + \phi_1) \\
	\end{pmatrix}
	\end{align}
 with $v_\perp = \sqrt{v_c^2 \cos^2 \phi + v_a^2 \sin^2 \phi}$, 
 $\phi_1 = \tan^{-1} (\tan \phi \frac{v_a}{v_c})$, the cyclotron energy $\omega_c = \sqrt{2 e B v_{\perp} v_b / \hbar}$, and a lowering operator $a \equiv \ell_B \left (i v_c (k_c +  B e r_\perp )  -   k_\perp v_\perp + k_\parallel (v_a^2-v_c^2) \sin \theta \cos \theta/v_{\perp} \right) / \sqrt{2 v_c v_\perp}$. \eqref{Hamiltonian_acdirection} can be diagonalized by the same procedure as specified in the text below \eqref{Hamiltonian_adirection}. 
 }

\begin{figure}[h]
	\centering
\NEW{
\includegraphics[width=7cm]{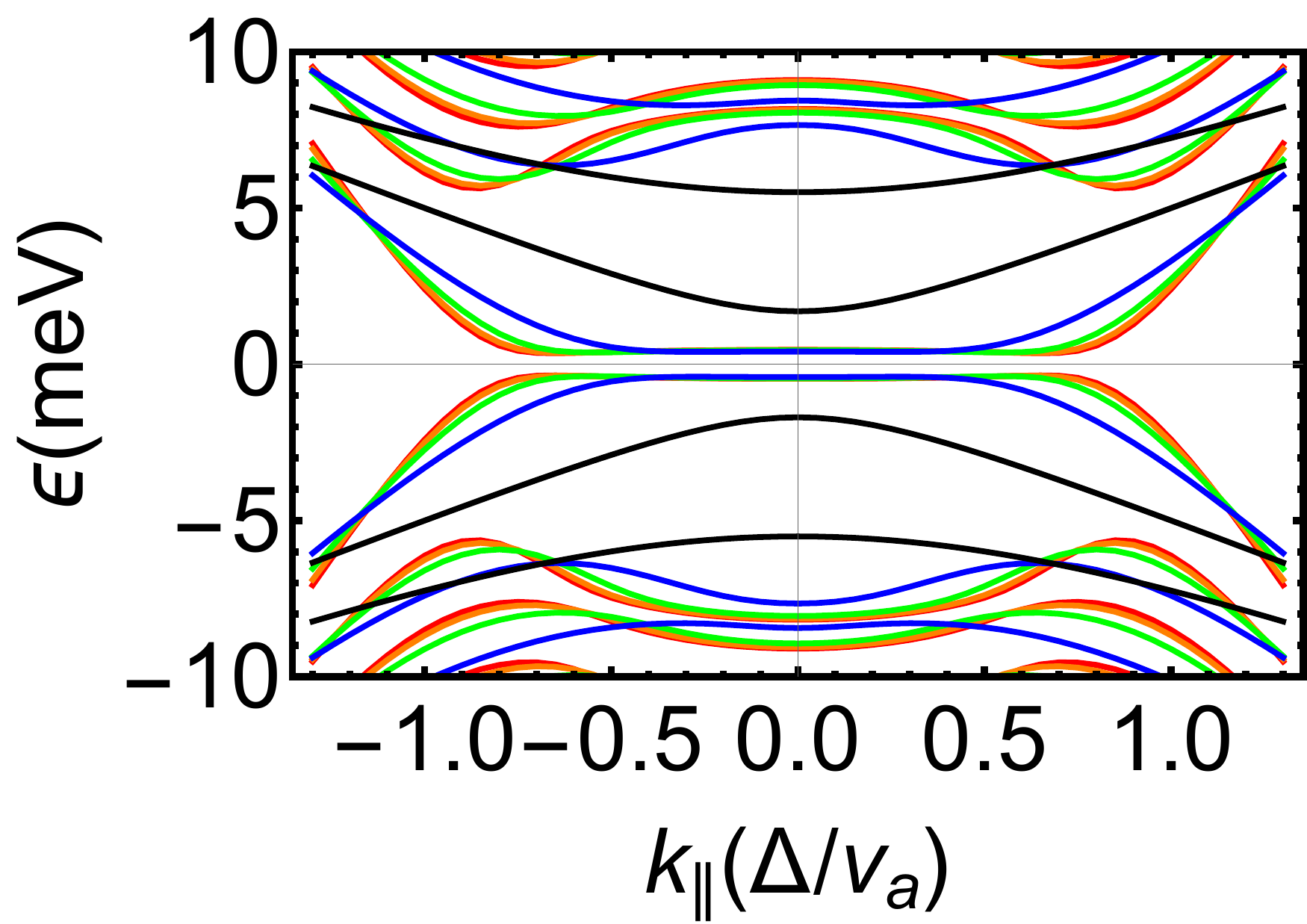}
	\caption{{\bf Dependence of the lowest Landau levels on the magnetic-field orientation $\phi$ in the $ac$ plane.} $\phi = 0$ (red), 5$^{\circ}$ (orange), 10$^{\circ}$ (green), 20$^{\circ}$ (blue), 90$^{\circ}$ (black). The magnitude of the magnetic field is $8$ T. We have also used the $g$ factor $g_B = 2$, $\Delta= 19.1$ meV, $v_b = 0.43 \times 10^5$ m/s, $v_a = 6.9 \times 10^5$ m/s, and $v_c = 1.7 \times 10^5$ m/s.}
	\label{SuppleFig:Bandstructureacdirection}
	}
\end{figure}

 \NEW{The band structures with different angle $\phi$ are plotted in Fig.~\ref{SuppleFig:Bandstructureacdirection}. $\phi = 0$ corresponds to a magnetic field along the $\hat{a}$ direction, as in Fig. \ref{fig:ac-rotation}a. For these plots, we have used the parameters of ZrTe$_5$ obtained from the quantum oscillation measurements~\cite{Wang2022} and $B$ = 8 T. Remarkably, flat bands are pronounced with magnetic-field orientation angles $\phi< 10^\circ$, and they are affected very slightly with $\phi$ in this regime. Note that compared with the band structures  (Fig.~\ref{SuppleFig:Bandstructure}) for the magnetic-field orientation in the torus plane, the angle dependence in the $ac$ rotation is weaker. Such a weak dependence is attributed to the smaller velocity anisotropy on the $ac$ plane ($v_a \approx 4 v_c$) than that on the torus plane ($v_a \approx 16 v_b$). These results are consistent with slower decay of the measured $\gamma$ with the rotation angle $\phi$ in the $ac$ plane (see Fig. \ref{fig:ac-rotation}b).
 In contrast, for larger angles $\phi > 20^\circ$, the lowest Landau levels acquire substantial dispersion and has a maximum velocity at $\phi = 90^\circ$ (i.e., $\vec{B}$ perpendicular to the torus plane) where the flat bands disappear.}

\subsection{Interplay between the puddle physics and nonlinearities}

Here we provide a qualitative discussion on the enhancement of nonlinear effects by large-scale inhomogeneities  arising from charged impurities. 

Our discussion starts from the  observation that the average electron density in our system is extremely small~\cite{Wang2022}, of the order of  $10^{16}$\,cm$^{-3}$, corresponding to less than $10^{-5}$ electrons per formula unit. Correspondingly, the screening length for charged impurities is very long and the unavoidable presence of charged impurities is expected to lead to large-scale inhomogeneities of the electronic system, the so-called electron-hole puddles. For a quantitative study of the formation of electron-hole puddles and their impact on transport, see, e.g., Ref.~\cite{Skinner2012,Borgwardt2016,Breunig2017}.

There are several mechanisms, how the formation of large-scale inhomogeneities of the electronic system, i.e., of puddles,
can strongly enhance pre-existing nonlinearities or create new types of nonlinear effects.
First, in a strongly inhomogeneous system, the electric field is not constant, creating locally regions of larger fields where nonlinear effects are enhanced. Thus, the average $\langle E^3 \rangle$ can become larger than $\langle E \rangle^3$. We believe, however, that this is unlikely to explain an enhancement of nonlinear effects by several orders of magnitude as we observe in our system.
A potentially much larger effect arises from the density dependence of the nonlinearity, 
$\gamma \sim 1/n^2$, see \eqref{Gamma3ZeroField} (the first term dominates over the second). Taking into account, that $n \sim \mu^2$, \eqref{DensityZeroField}, this implies that $\gamma \sim 1/\mu^4$ is a highly singular function of the local chemical potential $\mu(\vec r)$. In the presence of charged impurities and electron-hole puddles, $\mu(\vec r)$ is expected to vary strongly, even changing sign. This can therefore lead to a strong enhancement of nonlinear effects.

Unfortunately, a quantitative prediction (or even a rough estimate) of this nonlinear effect, which is non-perturbative both in interactions and disorder, turns out to be highly challenging and is beyond the scope of this work. 
The experimental observation of a very strong sample dependence of the nonlinearity $\gamma$ in combination with a 
relatively weak sample-dependence in $\gamma(B)/\gamma(B=0)$ is, however, consistent with this scenario. Figure 3 of the main text shows that the enhancement, $\gamma(B)/\gamma(B=0)$, is roughly given by a constant factor of $20$ even when absolute values of $\gamma$ vary by almost two orders of magnitude.

The formation of electron-hole puddles depends sensitively on both the average electron density and the precise density and ratio of positively and negatively charged impurities. Thus a very strong sample dependence has to be expected. At the same time, our theory for the {\em homogeneous} system predicts that the relative enhancement  $\gamma(B)/\gamma(B=0)$, which occurs when the system enters the flat-band regime deep in the quantum limit, depends only weakly on $n$ or $\mu$, see Fig.~5 of the main text. Thus,
it is plausible that $\gamma(B)/\gamma(B=0)$ shows much less sample-to-sample variations. 

We cannot exclude at the moment other sources of nonlinearities, e.g., related to intrinsic p-n junctions in regions where electron puddles and hole-puddles come close to each other. It is, however, completely unclear how such effects can explain the pronounced $B$-dependence of $\gamma$ and, especially, the insensitivity of $\gamma(B)/\gamma(B=0)$ to sample-to-sample variations.